\def\paperversion{tr}
\def\anonymoussubmission{1} 
\newif\ifsinglecolumn\singlecolumnfalse
\newif\ifwidemargins\widemarginsfalse
\newif\ifwarning\warningfalse
\newif\ifshowcomments\showcommentsfalse
\newif\ifblinded\blindedfalse
\newif\ifshowlinenums\showlinenumsfalse
\newif\ifreport\reportfalse
\newif\ifcopyrightspace\copyrightspacefalse
\newif\ifacknowledgments\acknowledgmentsfalse
\newif\ifshowpagenumbers\showpagenumberstrue
\newif\iffinalformat\finalformatfalse
\newif\ifweb\webfalse
\newif\ifexternalize\externalizetrue

\ifx\anonymoussubmission\undefined
\else
  \if\anonymoussubmission 1
    \blindedtrue
  \else
    \blindedfalse
  \fi
\fi
\IfFileExists{.blinded}{\blindedtrue}

\ifx\paperversion\xxxxundefined
\PackageError{paperversions}{*** No valid document version was specified.
Macro paperversions must be defined as one of (markup, draft,
local, submission, final, web, tr, trdraft)}
\fi

\def\xxversion{\csname xx\paperversion\endcsname}
\newif\ifsawversion\sawversionfalse

\ifcase\xxversion\relax
    \widemarginstrue
    \singlecolumntrue
    \warningtrue
    \showlinenumstrue
    \sawversiontrue
    \blindedfalse
\or 
    \warningtrue
    \showlinenumsfalse
    \showcommentstrue
    \sawversiontrue
\or 
    \warningtrue
    \showcommentsfalse
    \blindedfalse
    \sawversiontrue
    \acknowledgmentstrue
    \showlinenumstrue
\or 
    \sawversiontrue
    \showlinenumstrue
\or 
    \blindedfalse
    \sawversiontrue
    \copyrightspacetrue
    \acknowledgmentstrue
    \showpagenumbersfalse
    \finalformattrue
\or 
    \singlecolumntrue
    \blindedfalse
    \sawversiontrue
    \reporttrue
    \acknowledgmentstrue
    \webtrue
\or 
    \blindedfalse
    \singlecolumntrue
    \showcommentstrue
    \sawversiontrue
    \reporttrue
    \showlinenumstrue
    \warningtrue
\or 
    \blindedfalse
    \showcommentstrue
    \copyrightspacetrue
    \acknowledgmentstrue
    \sawversiontrue
    \finalformattrue
    \showlinenumstrue
    \warningtrue
\or 
    \blindedfalse
    \sawversiontrue
    \copyrightspacetrue
    \acknowledgmentstrue
    \finalformattrue
    \webtrue
\or 
    \singlecolumntrue
    \blindedtrue
    \acknowledgmentsfalse
    \sawversiontrue
    \reporttrue
    \webtrue
\or 
    \blindedfalse
    \sawversiontrue
    \copyrightspacetrue
    \acknowledgmentstrue
    \finalformattrue
    \externalizefalse
    \webtrue
\fi

\ifsawversion
\else
\fi

\let\xxversion=\undefined

\ifreport
    \documentclass[letterpaper,onecolumn,10pt]{article}
\else
    \iffinalformat
        \ifweb
            \documentclass[letterpaper,twocolumn,10pt]{article}
        \else
            \documentclass[letterpaper,twocolumn,10pt]{article}
        \fi
    \else
        \documentclass[conference]{IEEEtran}
    \fi
\fi

\ifreport
\title{Charlotte: Composable Authenticated Distributed Data Structures \\ \textit{Technical Report}}
\else
\title{Charlotte: A Web of Composable Authenticated Distributed Data Structures
  \ifblinded
       \vspace{-18mm}
  \fi
}
\fi

\ifblinded
\else
  \ifreport
   \author{
     {\begin{tabular}{c}
Isaac Sheff \\
Cornell University\\
\url{isheff@cs.cornell.edu}
\end{tabular}}
\and
{\begin{tabular}{c}
Xinwen Wang\\
Cornell University\\
\url{xinwen@cs.cornell.edu}
\end{tabular}}
\and
{\begin{tabular}{c}
Haobin Ni\\
Cornell University\\
\url{haobin@cs.cornell.edu}
\end{tabular}}
\and
{\begin{tabular}{c}
Robbert van Renesse\\
Cornell University\\
\url{rvr@cs.cornell.edu}
\end{tabular}}
\and
{\begin{tabular}{c c c c c}
Andrew C. Myers\\
Cornell University\\
\url{andru@cs.cornell.edu}
\end{tabular}}
   }
  \else
   \author{
     \IEEEauthorblockN{Isaac Sheff}
     \IEEEauthorblockA{
       {Cornell University}\\
       {\url{isheff@cs.cornell.edu}}}
     \and
     \IEEEauthorblockN{Xinwen Wang}
     \IEEEauthorblockA{
       {Cornell University}\\
       {\url{xinwen@cs.cornell.edu}}}
     \and
     \IEEEauthorblockN{Haobin Ni}
     \IEEEauthorblockA{
       {Cornell University}\\
       {\url{haobin@cs.cornell.edu}}}
     \and
       \IEEEauthorblockN{Robbert van Renesse}
       \IEEEauthorblockA{
       {Cornell University}\\
       {\url{rvr@cs.cornell.edu}}}
     \and
     \IEEEauthorblockN{Andrew C. Myers}
     \IEEEauthorblockA{
       {Cornell University}\\
       {\url{andru@cs.cornell.edu}}}
   }
 \fi

\fi
\ifreport
\else
\fi

\ifblinded
\else
  \acknowledgmentstrue
\fi
 
\usepackage{epsfig, times, graphicx, graphbox, color, eso-pic, tikz, multirow}
\usepackage{amsmath, amsthm, amssymb, stmaryrd, utf8math, ttquot, inconsolata, checkpagelimit}

\usepackage{hyperref}
\hypersetup{
	colorlinks,
	urlcolor=black,
	citecolor=black,
	filecolor=black,
	linkcolor=black
}
\ifshowcomments
    \usepackage[checkNoComments,margin]{aplcomments}
\else
    \usepackage[disabled]{aplcomments}
\fi

\usepackage{listings}
\usepackage{latex-listings-protobuf/protobuf/lang}
\usepackage{latex-listings-protobuf/protobuf/style}\SetProtoColorsBlueish

\usepackage{defs}

\let\crunchspace\relax
\ifx\crunchspace\undefined
\else
\usepackage{overrides}
\fi

\ifwarning
\AtBeginDocument{
 \AddToShipoutPicture*{\put(406,750){\it\color{red}{\fbox{\large Draft---please do not distribute}}}}
}
\fi
\pagelimit{13}
\AtEndDocument{\checkpagelimit}

\ifshowpagenumbers
\fi

\begin{document}
\hyphenation{time-stamp time-stamps}
\maketitle

\iffinalformat
  \thispagestyle{empty}
\else
\fi

\subsection*{Abstract}
We present \textit{Charlotte}, a framework for composable, 
 authenticated distributed data structures.
Charlotte data is stored in \textit{blocks} that reference each other
 by hash.
Together, all Charlotte blocks form a directed acyclic
 graph, the \textit{\blockweb}; all observers and applications
 use subgraphs of the \blockweb for their own data structures.
Unlike prior systems, Charlotte data structures are composable:
 applications and data structures can operate fully independently when
 possible, and share blocks when desired.
To support this composability, we define a language-independent format
 for Charlotte blocks and a network API for Charlotte servers.

An authenticated distributed data structure
guarantees that data is immutable and self-authenticating: data
 referenced will be unchanged when it is retrieved.
Charlotte extends these guarantees by allowing
 applications to plug in their own mechanisms for ensuring
 availability and integrity of data structures.
Unlike most traditional distributed systems, including distributed
 databases, blockchains, and distributed hash tables,
Charlotte supports heterogeneous trust: different observers may have 
 their own beliefs about who might fail, and how.
Despite heterogeneity of trust, Charlotte presents each observer
 with a consistent, available view of data.

We demonstrate the flexibility of Charlotte by implementing a variety
 of integrity mechanisms, including
 consensus and proof of work.
We study the power of disentangling availability and integrity
 mechanisms by building a variety of applications.
The results from these examples suggest that developers can use
 Charlotte to build flexible, fast, composable applications with
 strong guarantees.

\section{Introduction}
\label{sec:intro}

A variety of distributed systems obtain data integrity assurance by
 building distributed data structures in which data blocks are
 referenced using collision-resistant
 hashes~\cite{collision-resistance}, allowing easy verification that
 the correct data has been retrieved via a reference.
We call these \textit{\Adds} (\addss).
A particularly interesting example of an \adds is a blockchain,
 but there are other examples, such as distributed hash tables as
 in CFS~\cite{cfs}, distributed version control systems like
 Git~\cite{git}, and file distribution systems like
 BitTorrent~\cite{bittorrent}.
However, an \adds does not automatically possess all properties
 needed by blockchains and other applications.
An \adds might fail to ensure availability, because a reference to
 data does not guarantee it can be retrieved.
It might even fail to ensure integrity, because an \adds might be
 extended in inconsistent, contradictory ways---for example, multiple
 new \blocks could claim to be the 7th in some blockchain.

Therefore, an \adds commonly incorporates additional mechanisms to
 ensure availability and integrity in the presence of malicious
 adversaries.
Some systems rely on gossip and incentive schemes to ensure
 availability, and consensus or proof-of-work schemes to ensure
 integrity.
Blockchains like Bitcoin~\cite{bitcoin} and Ethereum~\cite{ethereum}
 lose integrity if the adversary controls a majority of the hash
 power, while Chord loses availability if an adversary controls enough
 consecutive nodes~\cite{chord}.

Importantly, all past \adds systems lack \textit{composability}:
 an application cannot use multiple \addss in a uniform way and
 obtain a composition of their guarantees.
\addss from different systems cannot intersect (share \blocks) or even
 reference each other.
Lack of composability makes it difficult for applications to
 atomically commit information to multiple \addss.
For instance, if blockchain \addss were composable, we could
 atomically commit a single \block to two cryptocurrency blockchains,
 instead of requiring trusted clearinghouses.

A core reason for this lack of composability is that each system has
 its own set of failure assumptions.
A user of \bitcoin or \ethereum, for example, must assume that at
 least half the hashpower is honest.\footnote{
   There is some evidence that users need even stronger
    assumptions~\cite{Eyal2018}.
 }
There is no mechanism for observers or applications to choose their
 own assumptions.

\begin{figure}
\centering
\begin{tabular}{cc}
\includegraphics[width=\ifreport0.5\else0.15\fi\textwidth,align=c]{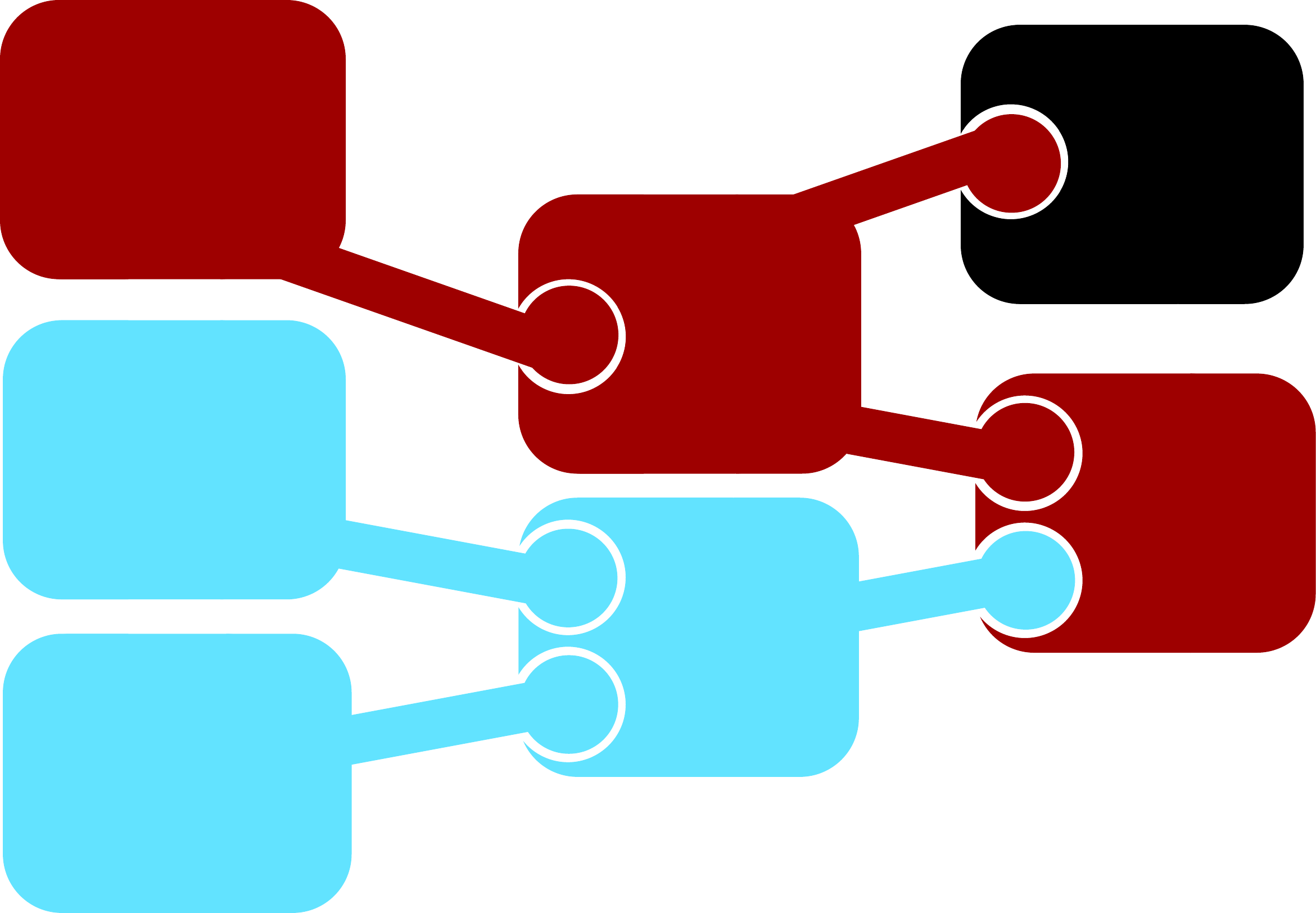} &
\begin{minipage}{2in}
\caption{Blocks are represented as rectangles.
References from one block to another are shown as circles.
The pale blue blocks form a tree, whereas
the darker red \blocks form a chain.
The rightmost red \block references a blue \block, so together the
 union of the red and blue \blocks forms a larger tree.
The black \block also references a red \block.
}
\label{fig:state}
\end{minipage}
\end{tabular}
\end{figure}

We address these limitations with \textit{Charlotte}, a decentralized
 framework for composable \adds with well-defined availability and
 integrity properties.
Together, these \addss form the \textit{\blockweb}, an authenticated
 directed acyclic graph (DAG)~\cite{Martel2001} of all Charlotte data,
 which is divided into \textit{\blocks} that reference each other by
 hash.
Charlotte distills \addss down to their essentials, allowing it to serve
as a common framework for building a wide variety of \addss in a
composable manner, as illustrated by
\autoref{fig:state}.


Within the \blockweb, different applications can construct any acyclic
 data structure from blocks, including chains, trees, polytrees,
 multitrees, and skiplists.
Whereas blockchains enforce a total ordering on all data, the
 \blockweb requires ordering only when one \block references another.
Unnecessary ordering is an enormous drain on performance; indeed,
 it arguably consumes almost all of traditional blockchains'
 resources.
Charlotte applications can create an ordering on \blocks, but
 \blocks are by default only partially ordered.

In Charlotte, each server stores whichever \blocks it wishes.
Most servers will want \blocks relevant to applications they're
 running, but some may provide storage or ordering as a service for
 sufficiently trusting clients.

Charlotte users can set their own (application-specific) failure
 assumptions.
The failure assumptions of a user effectively filter the \blockweb
 down to \blocks forming an \adds that remains available and
 consistent under all tolerable failures and adversarial attacks.
An observer whose failure assumptions are correct can, given
 the assumptions of a different correct observer, calculate the
 subgraph of the \blockweb they share.

A key novelty of Charlotte is its generality; it is not
 application-specific.
Unlike other systems that build DAGs of \blocks,
 Charlotte does not implement a
 cryptocurrency~\cite{iota,nano,avalanche,spectre,phantom},
 require a universal ``smart contract'' language for all
 applications~\cite{aeternity,alephium,qubic},
 have any distinguished ``main chain''~\cite{plasma, polkadot},
 or try to enforce the same integrity requirements across all
 \addss in the
 system~\cite{omniledger,Luu2016,rapidchain,rscoin,ethereum-sharding-phase-1}.

Instead, Charlotte distills \addss down their essentials, allowing it
to serve as a more general \textit{\adds framework}, in
 which each application can construct an \adds based on its own trust
 assumptions and guarantees, yet all of these heterogeneous \addss are
 part of the same \blockweb.
Indeed, existing \block-DAG systems can be recreated within Charlotte,
 gaining a degree of composability.
We have implemented example applications to demonstrate that Charlotte
 is flexible enough to simultaneously support a variety of
 applications, including Git-like distributed version control,
 timestamping, and blockchains based variously on agreement,
 consensus, and proof-of-work.
The shared framework even supports adding shared \blocks on multiple
 chains.

\subsection*{Contributions}
\begin{itemize}
\item Our mathematical model for \addss~(\autoref{sec:math})
    gives a general way to characterize \addss with diverse properties
        in terms of observers, a novel characterization of different
        failure tolerances for different
        participants, and
  a general way to compose \addss and their
         properties.
\item Charlotte provides an extensible type system for \blocks, and a
       standard API for communicating them~(\autoref{sec:api}).
\item Example applications show the benefits of using
       the Charlotte model~(\autoref{sec:use}).
\item We generalize blockchains in the Charlotte model, including a
      technique for separating availability and integrity duties onto
      separate services and a general model of linearizable transactions
      on distributed objects~(\autoref{sec:blockchainsstructures}).
\item We have implemented a prototype of Charlotte along with
      proof-of-concept implementations of various applications
      that demonstrate its expressiveness and ability to compose
      \addss (\autoref{sec:impl}).
\item Performance measurements show 
      that Charlotte's performance overheads are
      reasonable~(\autoref{sec:evaluation}).
\item Analysis of real usage data shows that Charlotte's added
       concurrency offers a large speed advantage over traditional
       blockchain techniques~(\autoref{sec:bitcoinanalysis}).
\end{itemize}

\section{Overview}
\label{sec:overview}
\vspace{-1mm}

\subsection{Blocks}
\label{sec:overviewblocks}
In Charlotte, \blocks are the smallest unit of data, so clients don't
 fetch ``\block headers,'' or other partial \blocks~\cite{ethereum}.
Therefore, Charlotte applications ideally use small \blocks.
For instance, to build something like \ethereum in Charlotte, it would
 be sensible to create the Merkle tree~\cite{merkle-trees} structure
 found within each \ethereum \block out of many small Charlotte
 \blocks.
This makes it easier to divide up storage duties and to fetch and
 reference specific data.

\subsection{Attestations}
\label{sec:overviewattestations}
Some \blocks are \textit{\attestations}:
 they prove that an \adds satisfies properties beyond those inherent
 to a DAG of immutable blocks.
For instance, if a server signs an \attestation stating that it will
 store and make available a specific \block, it means the
 \block will be available as long as that server functions correctly.
Such an \attestation functions as a kind of proof premised on the
trustworthiness of the signing server.
Attestations about the same \blocks naturally compose: all
 properties of all \attestations hold when all conditions are met.

All \attestation types are \textit{pluggable}: Charlotte servers can
 define their own subtypes, which prove nothing to observers who do
 not understand them.
Charlotte is extremely flexible: application-defined \attestation
 types can represent different consensus mechanisms
 (from Paxos to Nakamoto), different \adds types, and
 different availability strategies.
Although \attestations can express a wide variety of properties about
 an \adds, we divide them into two subtypes:
 \textit{\availabilityattestations} and
 \textit{\integrityattestations}.

\subsection{Availability Attestations}
\label{sec:overviewavailability}
Availability \attestations prove that \blocks will be
 available under certain conditions.
One example of an \availabilityattestation would be a signed statement
 from a server promising that a given \block will be available
 as long as the signing server is functioning correctly.
We call servers that issue \availabilityattestations
 \textit{Wilbur servers}.\footnote{after the \textit{Charlotte's Web} character whose objective is
    to stay alive.}
Attestations may make more complex promises.
For example, proofs of retrievability~\cite{Bowers2009} might be used
 as \availabilityattestations.
Availability \attestations are not limited to promises to store
 forever: they might specify any conditions, including time limits or
 other conditions under which the \block is no longer needed.
Availability \attestations generalize features found in many existing
 distributed data systems:
\begin{itemize}
\item In BitTorrent, a seeder tells a tracker that it can provide
       certain files to leechers.
\item Many databases inform clients that their
       transaction has been recorded by a specified set of replicas.
\item In existing blockchains, clients wait for responses
       from many full nodes, to be sure their transaction is
       ``available.''
\end{itemize}
\label{sec:wilbur}


\subsection{Integrity Attestations}
\label{sec:overviewintegrity}
An \adds often requires some kind of permission to add a \block
 to its state.
For example, a blockchain typically requires that some set of servers
 (``miners'') decide that a particular \block uniquely occupies a given
 height in the chain.
Integrity \attestations determine which \blocks belong in
 which \addss.
For instance, servers maintaining a blockchain might issue an
 \integrityattestation stating that a given block belongs on the chain
 at a specific height; the server promises not to issue any
 \integrityattestation indicating that a different block belongs on
 the chain at that height.
Timestamps are another \integrityattestation type: they
 define an \adds consisting of all \blocks a specific server
 claims existed before a specific time.
We call servers that issue \integrityattestations
 \textit{Fern servers}.\footnote
{after the \textit{Charlotte's Web} character who decides which
  piglets belong.}

\label{sec:fern}
Fern servers generalize ordering or consensus services.
In blockchain terminology~\cite{bitcoin}, they correspond to
 ``miners,'' which select the \blocks belonging on the chain.


\smallskip

\begin{figure}
\centering
\includegraphics[width=\ifreport0.8\else0.45\fi\textwidth]{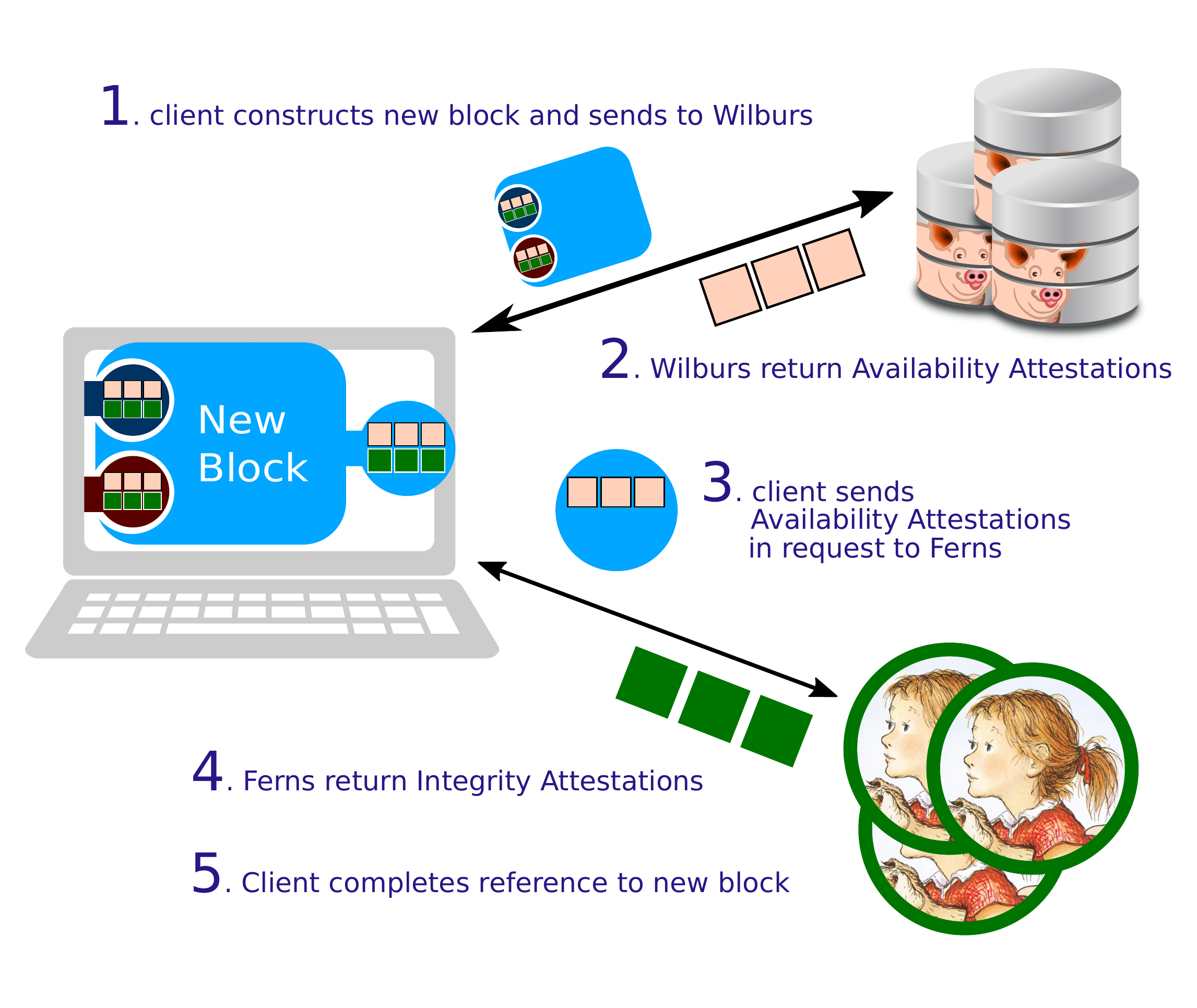}
\vspace{-2em}
\caption{Life of a \block.
A client mints a new block, and wants to add it to an \adds.
The block, as drawn, includes two references to other \blocks.
The client acquires \availabilityattestations from Wilbur servers, and
 \integrityattestations from Fern servers.
Then it can create a reference (drawn as a circle) to the block, so
 anyone observing the reference knows the block is in the \adds.}
\label{fig:life-of-a-block}
\end{figure}

\subsection{Life of a Block}
\label{sec:life-of-a-block}
\autoref{fig:life-of-a-block} illustrates one possible process for
adding a new block to an \adds.
A client first mints a \block, including data and references to other
 \blocks.
To ensure the \block remains available, the client sends it to
Wilbur servers, which store it and return
availability attestations, demonstrating the
availability of the block.

The client then submits a reference to the \block to a
 collection of \textit{Fern} servers, which maintain the integrity of
 the \adds.
Since Fern servers may not want to permanently add a \block to their
 \adds if that \block is going to become unavailable, the client may
 also send \availabilityattestations.
Fern servers return \textit{\integrityattestations}, that, in
 effect, demonstrate the integrity of the statement ``this \block is
 in this \adds.''

The client includes all of these \attestations in references to the
 \block, so that whenever an observer sees a reference to the \block,
 they know how available it is, and what \addss it belongs to.
Over time, more \attestations may be issued, so a \block can
 become more available or join more \addss, with greater integrity.

Charlotte is flexible: applications can optimize this process by
 co-locating services, forwarding attestations directly between
 servers, etc.

\subsection{Observers}
\label{sec:observers}

We characterize an \textit{observer} in a distributed system as an
 entity with a set of assumptions concerning the possible ways that
 the system can fail.
Note that failure types include both Crash and
 Byzantine~\cite{Lamport82}.
Given a set of assumptions about who can fail and how, and the desired
 integrity properties of each \adds, each observer may choose
 to ignore any portions of the \blockweb that lack
 adequate \attestations.
What remains is the observer's \textit{view} of the \adds: the set of
 blocks it believes are available and part of the state of the \adds.

Each observer's view of an \adds is guaranteed to remain
 available and to uphold any integrity properties the observer has chosen
 so long as the observer's failure assumptions hold.
Further, portions of the \blockweb that feature \attestations
 satisfying two observers are guaranteed to remain in both observers'
 views, once both have observed all the relevant blocks.
Of course, in practice, servers take time to download relevant
 \blocks, and in an asynchronous system there is no bound on the time
 this may take.

\subsection{Example Applications}
\label{sec:overview-applications}
\subsubsection*{Blockchains}
\label{sec:blockchains1}
Charlotte can easily represent blockchains---not only linear chains,
 but also more intricate sharded or DAG-based
 structures~\cite{Martel2001}.
Existing blockchain systems already effectively provide integrity and
 \availabilityattestations, phrased as proofs of work, proofs of
 stake, etc.
Charlotte makes these proofs more explicit, without limiting the \attestation
 types an application can use.
As a result, multiple chains can share a \block, if \attestations
 required for each all refer to the same \block.
By providing a framework in which applications can interact, but
 without prescribing a rigid data structure, Charlotte allows far more
 concurrency than monolithic chains like \ethereum that 
 totally order all \blocks into a single chain~\cite{ethereum}.
This flexibility is a natural realization of the database community's
 decades-old ideal of imposing a ``least ordering''~\cite{Bernstein}.

\subsubsection*{Distributed Version Control}
\label{sec:git}
Charlotte is also a natural framework for 
 applications like Git~\cite{git}.
Each Git \textit{commit} is a \block referencing zero or more parent
 commits.
A commit with multiple parents is a \textit{merge}, and a
 commit with no parents is a \textit{root}.
Each Git server stores and makes some commit \blocks available, and can
 communicate this fact with \availabilityattestations.
A Git server can also maintain \textit{branches}, which associate a
 branch name (a string) with a chain of commits.
When a server announces that it is making a new commit the head of a
 branch, it issues an \integrityattestation stating that the commit is
 part of the branch.

\subsubsection*{Public-Key Infrastructure}
\label{sec:pki}
Public Key Infrastructure (PKI) systems are almost always \addss.
Key endorsements are essentially \integrityattestations,
defining \addss such as the certificate trees used to
 secure HTTPS~\cite{x509} and the web of trust
 used to secure PGP~\cite{rfc4880}.
Keys and certificates can be retrieved by hash from dedicated
 storage servers such as
 PGP's keyservers~\cite{ldappgp,hkpthesis,hkprfc}, corresponding to
 Wilbur servers.
PKIs such as ClaimChain~\cite{Kulynych2018} already attest to
 and rely upon data structure properties, e.g., total ordering in
 chains.

\begin{figure}
\centering
\begin{tabular}{cc}
\includegraphics[align=c,height=10em]{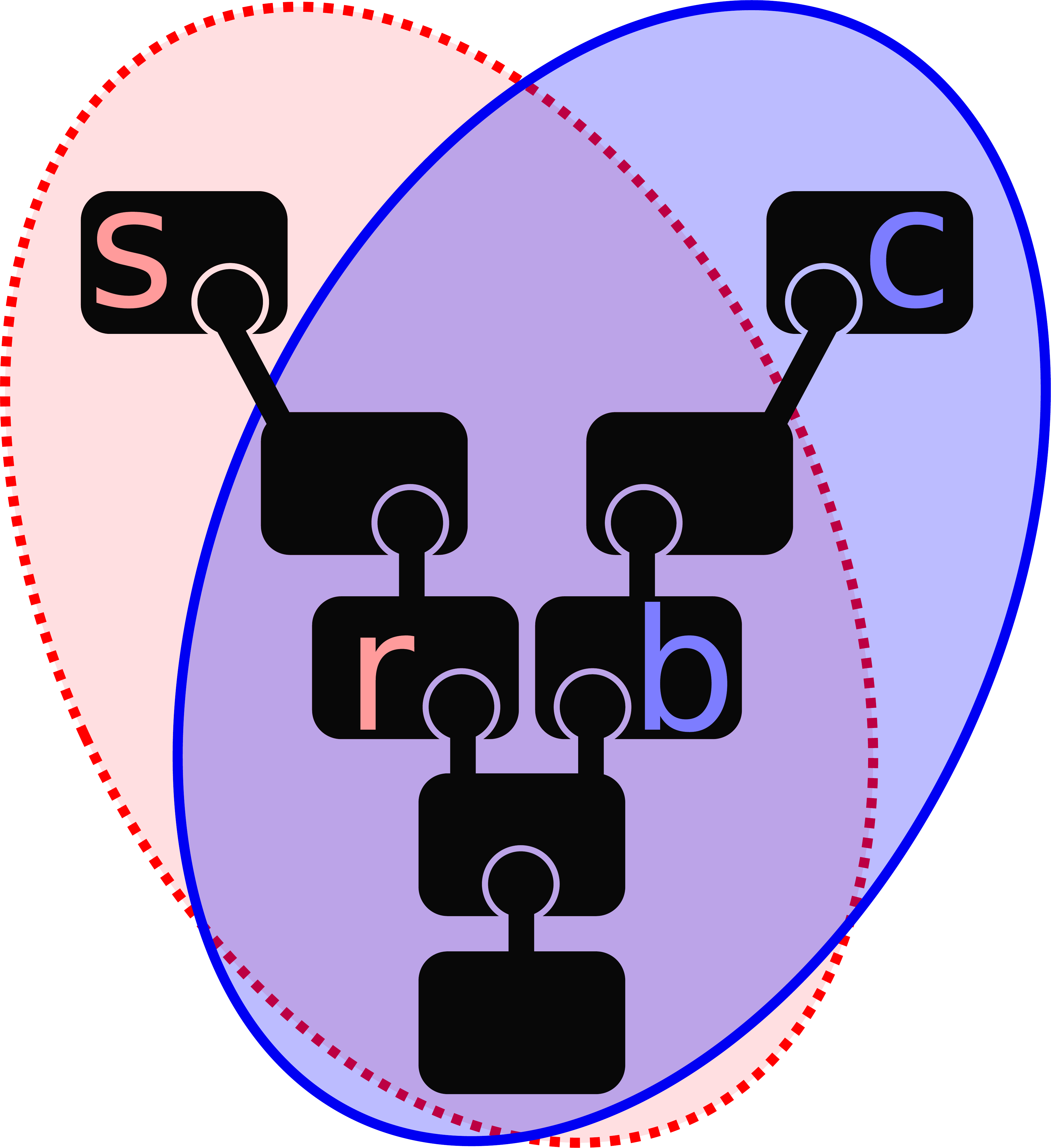} &
\begin{minipage}{2in}
\caption{A blockchain \adds with branches of length $< 3$.
         All of the \blocks are present in the \blockweb.
         The \red{red dotted oval} and the \blue{blue solid oval}
          represent two possible \textit{alternative} states of the
          \adds.
         }
\label{fig:chain-diverge}
\end{minipage}
\end{tabular}
\vspace{-2em}
\end{figure}

\section{Modeling ADDSs Formally}
\label{sec:math}
Different, possibly overlapping, portions of the \blockweb
 represent \addss of interest to individual applications.
We now explore Charlotte's unique ability to allow different
\addss to interoperate.

As a running example, consider a simple \adds $R$ representing a single,
 write-once slot managed by one server.
It can either be empty, or occupied by one unchanging block.

\subsection{States}
\label{sec:structures}

A \emph{state} is a set of \blocks, and an \emph{\adds} is
a set of possible states.
%
For instance, the \bitcoin blockchain is an \adds.
Every \block (other than the origin) in every state
 features a proof-of-work.
A \bitcoin state can have an arbitrarily long main chain, and
 shorter branches.
The \bitcoin~\adds consists of all such possible states.

In our single-slot example, each state of $R$ is either empty, or
 features exactly two blocks: the block occupying the slot,
 along with an \integrityattestation signed by the server,
 referencing that block.
We call an \integrityattestation in such a state $i_x$, where $x$ is
 the other block in the state.

\subsection{Observers and Adversaries}
\label{sec:observersBeliefsAdversaries}

Observers represent principals who use the system.
An observer receives \blocks from servers and in so doing learns about
 the current and future states of \addss in the system.
Observers may correspond (but are not limited) to servers, clients, or
 even people.
Formally, an observer is an agent that \textit{observes} an ordered
 sequence of \blocks from the \blockweb.
On an asynchronous network, different observers may see different
 blocks in different orders.

Observers define their own failure assumptions, such as who they
 believe might crash or lie.
These assumptions, combined with evidence, in the form of \blocks they
 have observed so far, induce an observer's \textit{belief}: what they
 think is true about the \blockweb now and what is (still) possible in
 the future.

The failure-tolerance properties of any distributed system are
 relative to assumptions about possible failures, including
 actions taken by adversaries.
Charlotte makes these assumptions explicit for each observer.
An observer who makes incorrect assumptions may not observe the
 properties they expect of some \addss.
For instance, if more servers are Byzantine than the observer thought
 possible, data they believed would remain available might not.
Alternatively, data structures might lose integrity, such as when two
 different \blocks both appear to occupy the same height on a chain.

\begin{figure}
  \centering
\includegraphics[width=\ifreport0.8\else0.47\fi\textwidth]{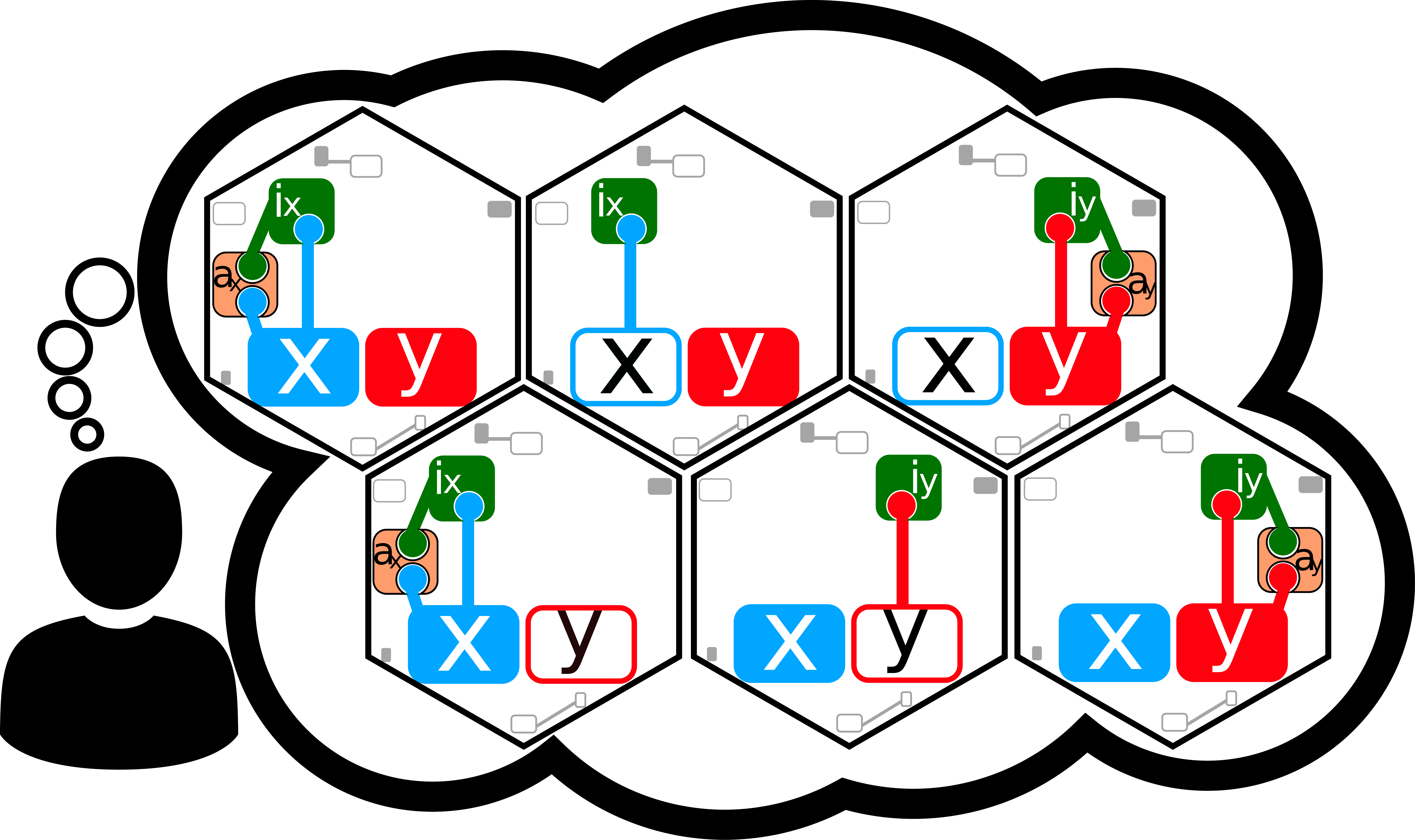}
  \vspace{-0.5em}
  \caption{
An \textit{observer} holds a \textit{belief}, which is a set of
 \textit{universes}.
Here we've drawn some universes as hexagons.
Each universe $U$ shown contains blocks in \exist U,
 with blocks in \avail U filled in.
}
 	\label{fig:observer}
\end{figure}

We characterize a belief $\alpha$ as a set of possible
 \textit{universes}.
This set bounds the believed powers of the adversary: the observer
 assumes this set includes all possible universes that might occur
 under the influence of the adversary.
\autoref{fig:observer} illustrates an observer holding a belief, and
 some of the universes in that belief.

Each observer has an \textit{initial belief}:
 the belief it holds before it observes any \blocks.
For example, an observer who trusts one Fern server to maintain the
 single-slot \adds $R$ does not have any universes in its initial
 belief in which that server has issued two \integrityattestations for
 different blocks.
This belief encodes the observer's assumption that the server's failure isn't
 tolerable.
The observer in \autoref{fig:observer} has such a belief: no
 universe features two integrity attestations for $R$
 (shown as green squares labeled $i_x$ or $i_y$).

In a traditional failure-tolerant system, an observer
 usually assumes that no more than $f$ participants will fail in some
 specific way (e.g., crash failures or Byzantine failures). 
We model such an observer's initial belief as the set of all
 universes in which no more than $f$ participants exhibit failure
 behaviors (in the form of blocks issued).

\subsection{Formalizing Universes}

We propose a general model for universes that places few limits on the
 details or assumptions universes can encode.
Our model of a universe $U$ has the following components, which 
 suffice for all examples in the paper.

 \medskip
 \noindent
 \fbox{
 \parbox{\columnwidth -1em}{
\begin{enumerate}
\item A set of \blocks that can \textit{exist}, written $\exist U$.
      These are the \blocks that either have already been observed or
       ever \textit{can} be observed by any observer.
\item A \textit{strict partial order} \observebefore U on \exist U.
      Every observer is assumed to observe \blocks in an order
       consistent with the universal partial order \observebefore U.
\item The set of \blocks that \textit{are available}, written
       \avail U.
      These are the blocks that can be retrieved from some server.
      Any available block must also exist:
       $\avail U \subseteq \exist U$.
\end{enumerate}
}}
 \medskip


The set \exist U constrains the blocks any observer will observe.
It does not model time: an observer's initial belief contains
 universes representing all possible futures, with all blocks
 that are possible in each.

Since we are modeling asynchronous systems, the model does
 not explicitly include the time when \blocks are observed,
 but the ordering \observebefore U constrains the times at which different
 observers can observe \blocks, implicitly capturing a
 temporal ordering on blocks.
This ordering is useful for blockchains like
 \bitcoin, where observers traditionally do not believe in any
 universe $U$ unless there is a \textit{main chain} in which
 each \block $b$ is ordered (by \observebefore U) before any equal-height
 \block with which $b$ does not share an ancestor fewer than security
 parameter $k$ (usually 6) \blocks away.
Further, the \textit{main chain} must forever outpace any other
 branch.
In \autoref{fig:chain-diverge}, this belief (with $k = 3$) implies
 that if a \bitcoin observer believes in a universe $U$ in which both
 {\blocks} \red s and \blue c exist, they must be ordered by
 \observebefore U. 
If \bitcoin's security assumptions are correct,
 any two observers must see \red s and \blue c in the same order.

We make the simplifying assumption in each of our example applications
 that the only availability of interest is permanent: we want
 to characterize whether \blocks will forever be available.
Hence, the set $\avail U$ increases over time.
We leave more nuanced availability policies to future work.

\subsection{Updating Beliefs}

As an observer observes \blocks being created by Charlotte programs,
 it updates its beliefs by whittling down the set of universes it
 considers possible.
For instance, if an observer with belief $\alpha$ observes a \block
 $b$, clearly $b$ \textit{can exist}, so the observer refines its
 belief.
It creates a new belief $\alpha'$, filtering out universes in which $b$
is impossible:
\[
  \alpha^\prime = \cb{\tallpipe{U\ }{\ b\in \exist U \ \land \ U \in\alpha}}
\]
If the observer in \autoref{fig:observer} were to observe $i_x$,
it would update its belief, retaining only universes $U$ with $i_x \in U$.
Of the universes shown, only the leftmost three would remain.

An observer also refines its belief by observation order:
If an observer with belief $\alpha$ observes \blocks $B$ in total order
 $<_B$, then its new belief is:
\ICS{We can combine the last two lines in the array, but I swear it's
      easier to read if we don't}
\[\arraycolsep=1pt
\possible \alpha B {<_B} \triangleq
\cb{\!\!\tallpipe{U\ }{\ 
  \begin{array}{r l}
    &\forall b^\prime\observebefore Ub.\ 
        b^\prime \in B\ \land\ b^\prime <_B b\\
    \land &   B \subseteq \exist U \\
    \land &   U \in\alpha
  \end{array}
}\!\!}
\]

An observer making no assumptions believes in all possible universes.
It can only eliminate universes inconsistent with its observations:
those in which \blocks it has observed are impossible, or the
 order in which it has observed the \blocks is impossible.
However, most interesting observers have other assumptions.
For example, the observer in \autoref{fig:observer} trusts that only
 one \integrityattestation for \adds $R$ will be issued, so if it
 observes $i_x$ and removes all universes $U$ without
 $i_x \in \exist U$, then no universes with $i_y$ will remain.

As another example, when a Git observer observes a valid
 \integrityattestation for a \block $b$, it can eliminate all
 universes with valid \integrityattestations for
 \blocks that are not descendants or ancestors of $b$.
\ICS{assuming it trusts the integrity attestation signer, and we're
 only talking about later blocks on the same branch, but I'm not sure
 it helps to write these extra assumptions in here.}

\label{sec:design}


\subsection{Observer Calculations}
\label{sec:observer-calculations}
An observer with belief $\alpha$ knows a set of \blocks $B$ are
 \textit{available} if they're made available in all possible
 universes:
\[
  \forall U \in \alpha.\ B \subseteq \avail U
\]
For example, the observer in \autoref{fig:observer} trusts
 availability attest\-ations $a_x$ and $a_y$ (the orange squares):
 it does not believe in any universe where such attestations
 reference an unavailable block.

Likewise, an observer with belief $\alpha$ knows a state $S$ of an
 \adds $D$ is \textit{incontrovertible} if no \textit{conflicting}
 state $S^\prime$ can exist in any possible universe.
Two states conflict if they cannot be merged to form a valid state:
 observing one precludes ever observing the other:
\[
  \forall U \in \alpha, S^\prime \in D.\ 
  \p{S \cup S^\prime \in D}\ \lor\ \p{S^\prime \not\subseteq \exist U}
\]
For example, the observer in \autoref{fig:observer} trusts
 that only one \integrityattestation for \adds $R$ will be issued.
It does not believe in any universes with both $i_x$ and $i_y$
 (shown as green squares).
Therefore, if it observes $i_x$, it knows the state $\cb{i_x, x}$ is
 incontrovertible: no conflicting state (such as $\cb{i_y, y}$) exists
 in any universe in its belief.

The state of \adds $D$ that an observer with belief $\alpha$ sees as
 available and incontrovertible is therefore:
\ICS{We can combine the last two lines in the array, but I swear it's
      easier to read if we don't}
\[\arraycolsep=2pt
\begin{array}{l}
  \viewTwo \alpha D \triangleq
\\
  \bigcup\cb{\tallpipe{\!\!S}{
    \begin{array}{r l}
    & \forall U \in \alpha, S^\prime \in D.\ 
                \p{S^\prime \subseteq S} 
                \lor
                \p{S^\prime \not\subseteq \exist U}\\
    \land & \forall U \in \alpha.\ S \subseteq \avail U\\
    \land & S \in D
    \end{array}
  }\!\!\!}
\end{array}
\]
We call this the observer's  \emph{view} of the \adds:
 Charlotte's natural notion of the ``current state.''
So long as an observer's assumptions are correct, new observations can
 only cause its view to grow.
For example, if the observer in \autoref{fig:observer} observes both 
 $a_x$ and $i_x$, then it believes the state $\cb{i_x, x} \in R$ is
 available and incontrovertible.
Its view of the single-slot \adds $R$ features $x$ occupying the slot,
 and so long as its assumptions are correct, this will never change.


As another example, suppose a blockchain uses a simple agreement
 algorithm: a quorum of servers must attest to a \block being at a
 specific height.
States consist of a chain of \blocks, each with \integrityattestations
 from a quorum.
An observer's view will not include any \blocks lacking sufficient
 \attestations.
The observer assumes that no two \blocks with the same height both
 get a quorum of \attestations, so the chain it
 has viewed must be a prefix of the chain in any future view.

One observer can calculate what another observer's view of an \adds
 would be, if they see the same observations.
When two observers communicate, they can share blocks they've
 observed.
Because new observations can only cause a view to grow, this allows
 one observer to know (at least part of) another observer's view when
 they communicate.
This what we mean when we say views in Charlotte are
 \emph{consistent:} two observers can know what the other views in the
 same data structure, and so the state of a data structure can be, in 
 a sense, global.




\subsection{Composability}
\label{sec:composability}
Recall that a \textit{state} is a set of \blocks, and an
 \adds is a set of states~(\autoref{sec:structures}).
\addss in Charlotte have two natural notions of composition:
  \textit{union} $\p\addsunion$ and \textit{intersection} $\p\addsintersection$.

\subsubsection{Union}
\label{sec:union}
Intuitively, the union of two \addss $D$ and $D'$ is all the data in
 either \adds.
As states are sets of \blocks (\autoref{sec:structures}), their union
 is simply the traditional union of sets.
Thus, the union \adds is composed of unions of states:
\[
  D \addsunion D^\prime \triangleq \cb{\tallpipe{S \cup S^\prime}{S \in D\ \land\ S^\prime \in D^\prime}}
\]
As a result, given an observer's failure assumptions, its view of the
union of two \adds is simply the union of its views of the \addss:
\begin{theorem}
  \[\forall \alpha, D. \viewTwo \alpha {D \addsunion D^\prime} = \viewTwo \alpha D \cup \viewTwo \alpha {D^\prime}\]
\end{theorem}
\begin{proof}
Follows from the definitions of $\viewTwoName$ and $\addsunion$.
\end{proof}

For example, a Git branch~(\autoref{sec:git}) is a \adds
 maintained by one server.
A Git repository is the \textit{union} of many branches with the same
 root, on the same server.
Each branch \adds has properties, such as linearity, not
 necessarily shared by the repository as a whole.
However, the properties of all the \addss in a \textit{union}
 can be combined to create properties that hold of the whole.
For example, one server makes available all the \blocks in all the
 branches of a repository.
That means that the repository remains available so long as the server
 is correct.
See \autoref{sec:observer-calculations} for more details.

\subsubsection{Intersection}
\label{sec:intersection}
Intuitively, the intersection of two \addss $D$ and $D'$ is all the
 data that is in both $D$ and $D'$.
As states are sets of \blocks~(\autoref{sec:structures}), their
 intersection is simply the traditional intersection of states.
Thus, the intersection of \addss is composed of the
 intersections of states:
\[
  D \addsintersection D^\prime \triangleq \cb{\tallpipe{S \cap S^\prime}{S \in D\ \land\ S^\prime \in D^\prime}}
\]
As a result, given an observer's failure assumptions, its view of the
intersection of two \adds is simply the intersection of its views of the \addss:
\begin{theorem}
  \[\forall \alpha, D. \viewTwo \alpha {D \addsintersection D^\prime} = \viewTwo \alpha D \cap \viewTwo \alpha {D^\prime}\]
\end{theorem}
\begin{proof}
Follows from the definitions of $\viewTwoName$ and $\addsintersection$.
\end{proof}

For example, consider two blockchains, each serving as a ledger for a
 different crypto-currency.
The \blocks that are part of both chains represent transactions atomically committed
 to both ledgers.
These are the natural place to put \textit{cross-chain transactions}:
 trades involving both crypto-currencies.
Thus, the intersection of the two blockchains is the sequence of
 cross-chain transactions.

The intersection \adds shares the properties of all
 intersected \addss.
In our blockchain example, the cross-chain \blocks remain totally
 ordered by the \blockweb so long as either component blockchain
 remains totally ordered by the \blockweb
 (a traditional integrity property of blockchains).
Furthermore, cross-chain \blocks remain available so long as the \blocks
 of either component blockchain remain available.
See \autoref{sec:observer-calculations} for more details.

\subsection{Availability Attestation Semantics}
\label{sec:availabilityattestations}


Observers use \availabilityattestations to determine which \blocks they
 consider sufficiently available to be in \addss they care
 about~(\autoref{sec:overviewavailability}).
Formally, subtypes $\tau$ of \availabilityattestation (which is in turn
 a subtype of \blocks) have values that guarantee some \blocks are
 available in some universes.
To describe the guarantees offered by an availability attestation, we
give a type $\tau$ an interpretation $\bb{\tau}$ that is a
\textit{belief}: that is, a set of universes in which
\availabilityattestations of that type are
inviolate~(\autoref{sec:observersBeliefsAdversaries}).

\mathname{AliceProvides}
\mathname{aliceProvides}

For instance, consider the \availabilityattestation subtype
 $\tau_\AliceProvides$.
Values of this type are \blocks of the form $\aliceProvides\p b$
 (where $b$ is another \block).
Intuitively, each value states that \texttt{Alice}
 (a Wilbur server) promises to make the specified \block $b$ available
 forever.
Thus, all universes $U$ in $\tau_{AliceProvides}$ in which
 $aliceProvides\p b$ exists also have $b$ available:
\[\arraycolsep=2pt
\begin{array}{l}
  \bb{\tau_{AliceProvides}} \triangleq \\
  \cb{\!\tallpipe U {
      \forall b.\ 
      aliceProvides\p b \in \exist U
      \ \Rightarrow\ 
      b \in \avail U
      \!
    }
  }
\end{array}
\]

Defining \attestations this way makes it easy to define observers'
 beliefs based on \textit{whom} they trust.
For instance, if an observer believes a \block will be available only
 if it has observed appropriate \attestations of both type $\tau$
 \textit{and} type $\sigma$, we define that belief $\alpha$ as 
$
  \alpha = \bb\tau \cap \bb\sigma
$.

Likewise, a more trusting observer who believes a \block is
 available if it has observed appropriate \attestations of type $\tau$
 \textit{or} type $\sigma$ would believe
 $\alpha = \bb\tau \cup \bb\sigma$.
In this way, we can even build up quorums of \attestation types
(e.g., $(\bb{\tau_1} \cap \bb{\tau_2}) \cup (\bb{\tau_2} \cap \bb{\tau_3}) \cup (\bb{\tau_1} \cap \bb{\tau_3})$).

There are some restrictions on the semantics of an
 \availabilityattestation type.
Attestations must be \textit{monotonic}: adding more \attestations
 never proves weaker statements:
\[\arraycolsep=1pt
\begin{array}{r r c l c l}
  \forall U, V, W \in \bb\tau.\ & \exist U & \cup & \exist V & \subseteq & \exist W\ 
    \Rightarrow\\
                             & \avail U & \cup & \avail V & \subseteq & \avail W
\end{array}
\]

\subsection{Integrity Attestation Semantics}
\label{sec:integrityattestations}
Integrity \attestations~(\autoref{sec:overviewintegrity}) are issued
 by Fern servers~\p{\autoref{sec:fern}}, and represent proofs
  guaranteeing the \textit{non-existence} of other
  \integrityattestations, under certain circumstances.
\ACM{This seems a bit circular. Shouldn't they also say something
  about the nonexistence of blocks that aren't integrity attestations,
  too? If integrity attestations only prove things about other
  integrity attestations, what is the point?}
\ICSreply{
  Integrity attestations can be part of the states of an \adds, so you
   can use the non-existence of an integrity attestation to show the
   non-existence of a state, and proving non-existence of conflicting
   states is what it's all about.
}
While this definition may seem counter-intuitive, it generalizes the
 notion of \textit{conflict} or \textit{exclusivity} in \addss.
For example, in our single-slot \adds $R$, all the
 \integrityattestations found in any state of $R$ are mutually
 exclusive.
Since each (non-empty) state of $R$ contains an \integrityattestation,
 the existence of one \attestation disproves all conflicting states, 
 which puts the attestation, and the block it references, in the view
 of any observer with an appropriate belief.


Formally, a subtype $\tau$ of \integrityattestation has values that
 guarantee some other \blocks \textit{will not exist} in some
 universes.
 
Thus, we represent every \attestation type $\tau$ as a set of
 universes, essentially a
 \textit{belief}~(\autoref{sec:observersBeliefsAdversaries}) in that
 type.
To describe \integrityattestations' guarantees, we have a static
 semantics where types are identified with \textit{beliefs}, sets of
 universes in which \integrityattestations of that type are
 inviolate~(\autoref{sec:observersBeliefsAdversaries}).

\mathname{BobCommits}
\mathname{bobCommits}

For example, consider $\tau_\BobCommits$, a subtype of
 \integrityattestation with values that are blocks of the form
 $\bobCommits\p{b}$, which intuitively indicates that \texttt{Bob} (a
 Fern server) promises never to commit to any \block other than $b$.
These \integrityattestations are much like the ones used in our
 single-slot \adds $R$.

Thus, all universes $U$ in which $\bobCommits\p{b}\in\exists{U}$ don't
 feature $\bobCommits\p{c}$ for any $c \ne b$:
\[\arraycolsep=2pt
\begin{array}{l}
  \bb{\tau_{\BobCommits}} \triangleq \\
  \cb{\tallpipe U {
      \forall b,c.\ b \ne c
      \ \Rightarrow\ 
      \cb{\begin{array}{l}
            \bobCommits\p b,\\
            \bobCommits\p c
          \end{array}}
      \not\subseteq \exist U
  }}
\end{array}
\]

\integrityattestation types with these semantics make it easy to
 define observers' beliefs based on \textit{who} they trust.
For instance, an observer who believes $b$ is committed only after
 receiving an \attestation of type $\tau$ and an \attestation of type
 $\sigma$ would believe $\alpha = \bb\tau \cup \bb\sigma$.
Likewise, a more trusting observer who believes $b$ is committed after
 receiving an \attestation of either type $\tau$ or $\sigma$ would
 believe $\alpha = \bb\tau \cap \bb\sigma$.
\ifreport

\ICS{I've relegated this fact to the tech report, because I'm not sure
      it's absolutely necessary to say.}
It is also possible to combine integrity and \availabilityattestation
 types to define a belief.
An observer who trusts \attestations of type $\tau$ to commit \blocks,
 and \attestations of type $\rho$ to ensure their availability would
 believe:
$\gamma = \bb\tau \cap \bb\rho$.
\fi
In this way, we can even define quorums of trusted types.

\ICS{Technically, we could have <C where <C subseteq <B, and the
statement still holds, but it's much uglier looking, to the
point of potentially being confusing.}
The definition of $\possible {\bb\tau} B {<_B}$
 (from \autoref{sec:observersBeliefsAdversaries}) guarantees
 \integrityattestations are \textit{monotonic}: adding more
 \attestations never proves weaker statements:
\[
  C \subseteq B \Rightarrow
  \possible \tau B {<_B}
  \subseteq
  \possible \tau C {<_B}
\]

\ifreport
\ICS{On Matthew's advice, I've relegated this section to the tech
 report / appendix.
It consists of us saying we don't do a thing that we never claim to
 do, and then an implementation detail that most readers don't care
 about.}
\subsection{Implementation Limitations of Attestations}
\label{sec:limitationsofattestations}
Since programmers can define their own subtypes of integrity or
 \availabilityattestations, nothing prevents them from encoding
 availability guarantees in an \integrityattestation, or violating the
 \availabilityattestation monotonicity
 requirement~(\autoref{sec:availabilityattestations}).
Programmers who violate the system assumptions naturally lose 
 guarantees.

\ICS{It would be nice to condense this paragraph, but I'm not sure how.}
In our implementation, the only operational distinction between an
 \availabilityattestation and an \integrityattestation is in the
 \texttt{Reference} object.
When one \block references another, it can also reference relevant
 integrity and \availabilityattestations. 
However, whereas an included reference to an \integrityattestation is
 itself a \texttt{Reference} object, an included reference to an
 \availabilityattestation carries only a \texttt{Hash}.
This is because an \integrityattestation might need an
 \availabilityattestations to describe where to obtain the
 \integrityattestation.
However, the same is not true of an \availabilityattestation: it is
 pointless to send \availabilityattestation $b$ just
 to describe where to fetch \availabilityattestation $a$, since it is
 just as easy to send \availabilityattestation $a$ in the first place.

\fi

\section{Charlotte API}
\label{sec:api}

\ifreport
\begin{figure}
\centering
\begin{lstlisting}[language=protobuf3,style=protobuf]
message AnyWithReference {
 google.protobuf.Any any;
 Reference typeBlock;}
message Hash {
 oneof hashalgorithm_oneof
   { AnyWithReference any;
     bytes sha3; }}//technically unnecessary
message Reference {
 Hash hash;
 repeated Hash availabilityAttestations;
 repeated Reference integrityAttestations;}
message Block {
 oneof blocktype_oneof
  { AnyWithReference any;
    string protobuf; }}
\end{lstlisting}
\caption{Core Types of Charlotte: this (slightly simplified) proto3 code describes how \blocks, references to \blocks, and generic data are safely marshaled and unmarshaled in Charlotte. }
\label{fig:types}
\end{figure}
\fi

Charlotte is a set of protocols by which clients, Fern servers, and
 Wilbur servers interact.
Different servers can run different implementations of these
 protocols.
Our implementation of Charlotte~(\autoref{sec:impl}) uses
 gRPC~\cite{grpc}, a popular language-independent network service
 specification language, based on Protocol Buffers~\cite{protobufs}.
\ifreport
Hence, we use Protocol Buffer (protobuf) syntax to describe the
 Charlotte protocols.
\else
\fi

\ifreport
\autoref{fig:types} presents the  core types used by Charlotte
 protocols, using Protocol Buffer syntax.\footnote{
   For simplicity, our specifications omit the indices of the various
    fields.
   The actual source code is also slightly more complicated for
    extensibility~\cite{anonymized-charlotte-code}.
}
\fi
\begin{samepage}
Charlotte is built around these core types:
\begin{itemize}
  \item \tb{\texttt{Block}}: can contain any protobuf~\cite{protobufs} data
                    type, or the \block itself can be a protobuf type
                    definition.  \texttt{Attestation} is a subtype of
                    \texttt{Block}.
                   \ifreport
                   It can contain any protobuf~\cite{protobufs} data
                    type, and the \block itself can be a protobuf type
                    definition.
                   \fi
  \item \tb{\texttt{Hash}}: represents the hash of a \block.
  \item \tb{\texttt{Reference}}: is used by one \block to reference another;
                        it contains the \texttt{Hash} of the referenced \block,
                        along with zero or more references to \attestations
                         (\autoref{sec:overviewattestations}).
  \item \tb{\texttt{AnyWithReference}}: Anyone can add their own subtypes of
                                \texttt{Block}, \texttt{Hash}, or
                                \texttt{Attestation}, which any
                                server can safely marshal and unmarshal.
                               It contains a reference to the \block
                                where the type description can be
                                found (as
                                proto3~\cite{protobufs} source code),
                                and marshaled data.\footnote
                                { The proto3 \texttt{Any} type itself
                                  features a URL string meant to
                                  reference the type definition, but
                                  Charlotte uses a \block reference
                                  because it is self-verifying.}
\end{itemize}
In practice, we provide some useful example subtypes of \texttt{Hash}
 (e.g., \texttt{sha3}) and \texttt{Block} (e.g., \texttt{Attestation}).
\end{samepage}

\ifreport

\begin{figure}
\centering
\begin{lstlisting}[language=protobuf3,style=protobuf]
message SendBlocksResponse {
 string errorMessage;}
service CharlotteNode {
 rpc SendBlocks(stream Block)
     returns (stream SendBlocksResponse) {}}
\end{lstlisting}
\caption{All Charlotte servers implement the Charlotte\-Node service.}
\label{fig:charlottenodeservice}
\end{figure}

In our API, all Charlotte servers must implement the
 \texttt{SendBlocks} RPC (\autoref{fig:charlottenodeservice}),
 which takes in a stream of \blocks and can return a stream of
 responses that may contain error messages.
We define subtypes of \attestation for \textit{Availability} and
 \textit{Integrity}, and show how to construct and observer from
 quorums of types they trust~(\autoref{sec:availabilityattestations}
 and \autoref{sec:integrityattestations}).

\subsection{Wilbur}
\label{sec:wilbur}

\begin{figure}
\centering
\begin{lstlisting}[language=protobuf3,style=protobuf]
message AvailabilityPolicy {
 oneof availabilitypolicytype_oneof {
  AnyWithReference any; }
}
message RequestAttestationResponse {
 string errorMessage;
 Reference reference;
}
service Wilbur {
 rpc RequestAvailabilityAttestation(
      AvailabilityPolicy)
     returns (RequestAttestationResponse){}
}
\end{lstlisting}
\caption{Wilbur Service Specification.}
\label{fig:wilburservice}
\end{figure}

Wilbur servers host \blocks, providing \textit{availability}.
\ICS{Removed the bit actually describing at a high level what Wilbur 
      servers do, since we've said that multiple times before.
     Should I bring it back?}

In blockchain terminology~\cite{bitcoin}, Wilbur servers correspond to
 ``full nodes,'' which store \blocks on the chain.
In more traditional data store terminology, Wilbur servers are
 key--value stores for immutable data.
The Charlotte framework is intended to be used for building both kinds
 of systems.

In our API, Wilbur servers are Charlotte servers that include the
 \texttt{RequestAvailabilityAttestation}
 RPC~(\autoref{fig:wilburservice}), which accepts a description of the
 desired \attestation, and returns either an error message, or a
 reference to a relevant \availabilityattestation.

\subsection{Fern}
\label{sec:fern}

\begin{figure}
\centering
\begin{lstlisting}[language=protobuf3,style=protobuf]
message IntegrityPolicy {
 oneof integritypolicytype_oneof
  { AnyWithReference any; }
}
service Fern {
 rpc RequestIntegrityAttestation(
      IntegrityPolicy)
     returns (RequestAttestationResponse){}
}
\end{lstlisting}
\caption{Fern Service Specification.}
\label{fig:fernservice}
\end{figure}

Fern servers issue
 \textit{\integrityattestations}, which define the set of \blocks in a
 given \adds.
Among other things, \integrityattestations can be proofs-of-work, or
 records demonstrating some kind of consensus has been reached.
One simple type of \integrityattestation, found in our prototype, is a
 signed pledge not to attest to any other \block as belonging in a
 specific slot in an \adds.
Fern servers generalize ordering or consensus services.
In blockchain terminology~\cite{bitcoin}, Fern servers correspond to
 ``miners,'' which select the \blocks belonging on the chain.

In our API, Fern servers are Charlotte servers that include the
 \texttt{RequestIntegrityAttestation} RPC~(\autoref{fig:fernservice}),
 which accepts a description of the desired \attestation, and returns
 either an error message or a reference to a relevant
 \integrityattestation.

\subsection{Practices for Additional Properties}
\label{sec:practices}
\ICS{Better section title needed?}

In order to understand a reference object within a block
 (how available the referenced block is, and data structures it's in),
 an observer reads attestations referenced within the reference
 object.\footnote{
   We considered making references contain full copies of
    \attestations, but this made blocks large, and since many blocks
    may reference the same block (and attestations), blocks were full
    of redundant information.
 }
For example, without the content of the \availabilityattestations,
 it's not clear where to look to retrieve the referenced \block.
As a rule of thumb, before one server sends a \block to another, it
 should ensure the recipient has any \attestations or type
 \blocks\ifreport\else~(\autoref{sec:api})\fi\ referenced within that
 \block.
This ensures the recipient can, in a sense, fully understand the
 \blocks they receive.
In \autoref{fig:sendeverything}, for instance, when sending block $a$,
 the sender should be sure the recipient has received everything in
 the dashed rectangle.
Our example applications follow this practice.
It is possible, however, that for some applications, servers may be
 certain the recipient doesn't care about some attestations or type
 blocks, and therefore might leave those out.

\begin{figure}
\centering
\begin{tabular}{cc}
\hspace{-6mm}\includegraphics[align=c,width=\ifreport0.5\else0.25\fi\textwidth]{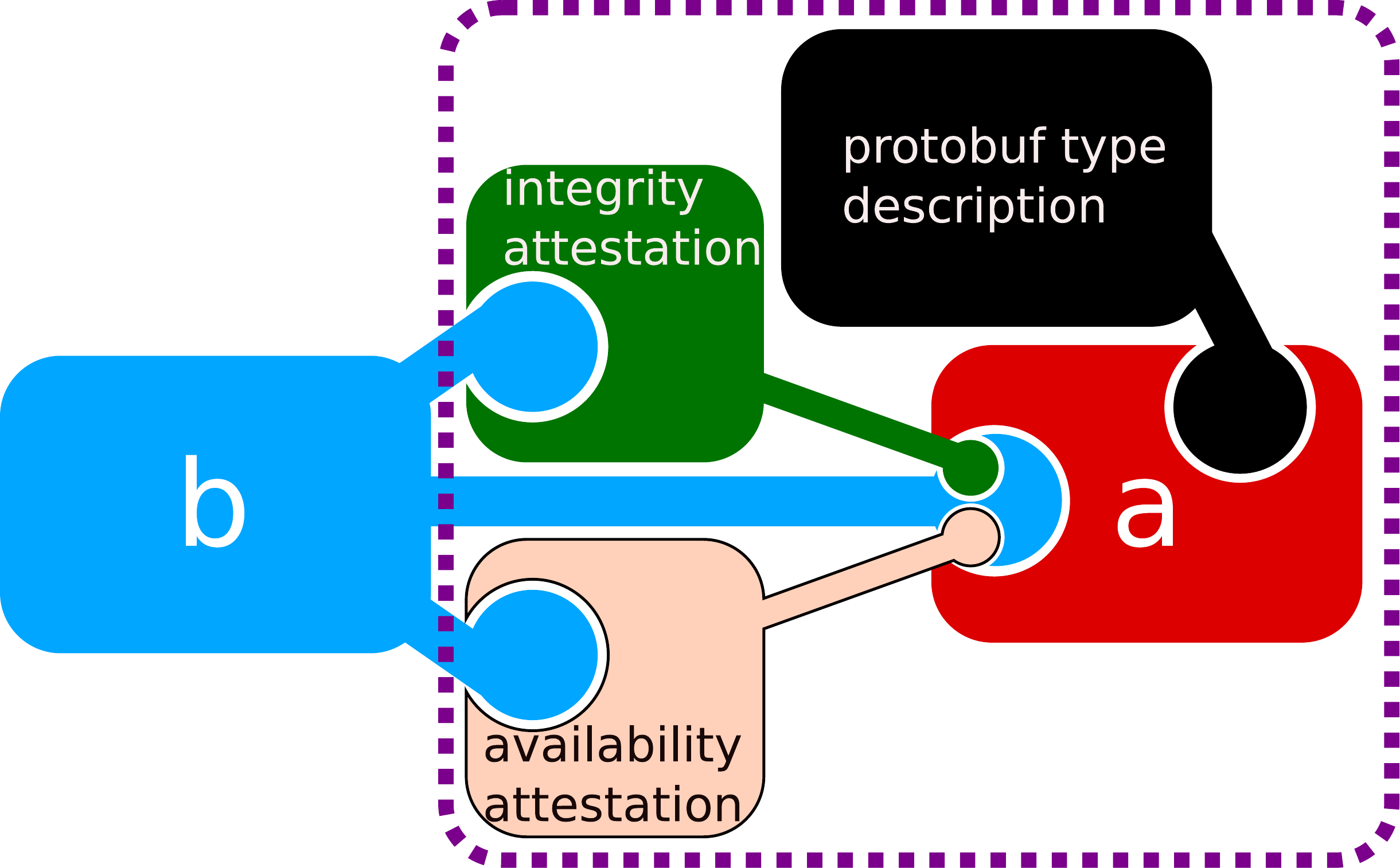}
\ICS{This figure can probably be redone better in TikZ.}
& \begin{minipage}{1.6in}
\caption{Block $a$ references \block $b$, and that reference carries
          \attestations.
         Block $a$ also references a type description block, for
          unmarshaling data in $a$.
         In general, when sending \block $a$ to a server or client,
          the sender should be sure the recipient has received
          \textit{all} the blocks in the dashed purple rectangle, so
          the recipient can fully understand \block $a$ and the
          properties of its references.}
\label{fig:sendeverything}
\end{minipage}
\end{tabular}
\end{figure}

When servers follow this practice, it's useful for
 \availabilityattestations to attest to groups of blocks likely to be
 requested together. 
In \autoref{fig:sendeverything}, for instance, and
 \availabilityattestation that attests to everything in the dashed
 rectangle would be more useful than just attesting to \block $a$.
Our example applications' \availabilityattestations are generally
 designed this way.

\ICS{The paragraph below could in principle be cut.
     It's not necessary to understand anything else in the paper.}
Availability failures can cause available states of \addss to become
 disconnected subgraphs (if the blocks that connect them are
 forgotten). 
To build an \adds that will always remain connected,
 \availabilityattestations that attest to a \block should also attest
 to the \availabilityattestations referenced within that \block.
Furthermore, whenever a \block $x$ references a \block $y$, and
 \block $y$ references \block $z$, if $y$ isn't at least as available
 as $z$, then $x$ should reference $z$ as well.
(Here, ``isn't at least as available'' means that the
 \availabilityattestations in references to $z$ guarantee $z$ will be
 available in some universe where the \availabilityattestations in
 references to $y$ do not guarantee $y$ will be available.)

\else

All Charlotte servers must implement the
 \texttt{SendBlocks} Remote Procedure Call (RPC),
 which takes in a stream of \blocks and can return a stream of
 responses that may contain error messages.
Additionally, Wilbur servers implement the 
 \texttt{RequestAvailabilityAttestation}
 RPC, and Fern servers implement the 
 \texttt{RequestIntegrityAttestation} RPC,
 each of which accept a description of the
 desired \attestation, and return either an error message, or a
 reference to a relevant \attestation.
Our full API is open source~\cite{anonymized-charlotte-code}.

We detail the gRPC spec in
 Appendix~\ref{sec:specificationsappendix}, and discuss some best
 practices for using the Charlotte API in
 Appendix~\ref{sec:practices}.
\fi

\section{Use Cases}
\label{sec:use}
\ICS{Better section title needed?}
In addition to the examples mentioned
 earlier~(\autoref{sec:overview-applications}), Charlotte is
 well-suited to a wide variety of applications.

\subsection{Verifiable Storage}
\label{sec:verifiablestorage}
\ICS{Is this paragraph worth having?}
Our Wilbur specification provides a common framework for verifiable
 storage.
Because \adds references include hashes, it is always
 possible to check that data retrieved was the data referenced.
Furthermore, \availabilityattestations
 (\autoref{sec:availabilityattestations}) are a natural framework for
 proofs of retrievability~~\cite{Bowers2009}.

\subsubsection{Queries}
\label{sec:wilburquery}
\ifreport
\begin{figure}
\centering
\begin{lstlisting}[language=protobuf3,style=protobuf]
message WilburQueryInput {
 oneof wilburquery_oneof {
  Reference reference = 1;
  Block fillInTheBlank = 2; }}
message WilburQueryResponse {
 string errorMessage = 1;
 repeated Block block = 2; }

service WilburQuery {
  rpc WilburQuery(WilburQueryInput)
      returns (WilburQueryResponse) {}}
\end{lstlisting}
\caption{WilburQuery Specification.}
\label{fig:wilburqueryservice}
\end{figure}
\fi

In addition to \texttt{SendBlocks} and
 \texttt{Request\-Availability\-Attestation}, Wilbur servers may
 offer other interfaces.
Application designers may wish to implement query systems for
 retrieving relevant \blocks.
We created one such example interface, the \texttt{WilburQuery}
 RPC\ifreport~(\autoref{fig:wilburqueryservice})\fi.
Given a \texttt{Hash} as input, \texttt{WilburQuery} returns the
 \block with that hash.
 \ifreport
If the server does not know of such a \block, our example
implementation waits until one arrives.
\fi

\texttt{WilburQuery} also provides a kind of fill-in-the-blank
 match:
If sent a \block with some fields missing, \texttt{WilburQuery} returns the
all stored \blocks that match the input \block in the provided
fields. For example, we might query for all \blocks with a field
 marking them as a member of a certain \adds.

\subsection{Timestamping}
\label{sec:use-timestamping}
Timestamps are a subtype of \integrityattestation.
We implemented a Signed Timestamp type, wherein the signer promises
 that they have seen specific hashes before a specific time.
Our timestamping Fern servers can use \textit{batching}: they wait for
 a specific (configurable) number of new requests to arrive before
 issuing a Timestamp \block referencing all of them.
In fact, since hash-based references represent a
 \textit{happens-before} relationship~\cite{Lamport78},
 timestamps are transitive:
 if timestamp $a$ references timestamp $b$, and $b$ references $c$,
 then $a$ effectively timestamps $c$ as well.

We recommend that batch Timestamp \blocks themselves should be
 submitted to other Timestamping Fern servers.
This allows the tangled web of Timestamp \blocks to very quickly stamp
 any \block with exponentially many timestamps, making them very
 high-integrity.

\subsection{Conflict-Free Replicated Data Types}
\label{sec:crdts}
Charlotte, and \addss in general, work well with CRDTs, especially
 Operation-Based Commutative Replicated Data Types
 (CmRDTs)~\cite{Shapiro2011}.
CmRDTs are replicated objects maintained by a group of servers.
Whenever a new operation originates at any server, all known
 operations on that object on that server are said to
 \textit{happen before} it.
Then the operation asynchronously propagates to all other servers.
Thus, the set of operations known to any particular server are only
 partially ordered.
The \textit{state} of a CmRDT object is a deterministic function of
 a set of known operations (and their partial order).
For example, a CmRDT implementation of an insert-only Set might
 feature the \texttt{insert} operation, and its state would be the set
 of all arguments to known \texttt{insert} operations.

In Charlotte, CmRDT operations can naturally be expressed as \blocks,
 with \textit{happens-before} relationships expressed as references.
Since references are by hash, it is impossible for an adversary to
 insert a cycle into the graph of operations.
The states of a CmRDT can be formally expressed as all possible sets
 of operations with all possible partial orderings.

Aside from whatever credentials one needs to authorize an operation,
 CmRDTs do not need \integrityattestations.
Observers need only consider the graph of known, valid operation
 \blocks with known ancestry.
They may, of course, choose to filter out \blocks they consider
 insufficiently available.
Availability \attestations are still useful.

The \blockweb as a whole is a CmRDT:
Its state is the DAG of all \blocks, and every \block is an operation
 adding itself to the state.
Other than the \blockweb itself, however, we have not implemented any
 interesting CRDTs yet.

\subsection{Composition}
\label{sec:composition}
Charlotte \addss are easy to
 compose~(\autoref{sec:composability}).
At the most basic level, \blocks in one \adds
 can reference \blocks in another.
For instance, a Timestamp server might maintain a chain of timestamp
 \blocks, which reference any other \blocks people want timestamped (Git
 commits, payments, documents, etc.).
A Git-style repository might reference earlier commits in another
 repository (either because one is a fork of the other, or one has
 merged in code from another) without having to copy all of the data
 onto both servers.
This would resemble Git's \emph{submodule} system~\cite{git-submodules}.
A blockchain could reference a Git commit as a smart contract, instead
 of hosting a separate copy of the code~\cite{ethereum}.
A single \block of data, stored on some highly available servers, could
 be referenced from a variety of torrent-style filesharing
 applications, git-style repositories, and blockchains, without
 unnecessary duplication.

At a high level, composability allows us to build high-integrity
 \addss out of low-integrity ones~(\autoref{sec:composability}).
For instance, the \blocks that appear in the intersection of two chains
 form a chain that can only fork if \textit{both} component chains
 fork.
Users may want to put especially important \blocks on many different
 chains, the way they want many different witnesses for important
 legal transactions.

Likewise, we can build low-integrity \addss out of higher
 integrity ones~(\autoref{sec:composability}).
If a set of blockchains each manage independent tokens, and sometimes
 share \blocks (for atomic trades of tokens), then together all the
 chains form a DAG.
If \textit{any} chain in the DAG is corrupted, then the supply of that
 token may not be conserved: the DAG as a whole is lower integrity
 than any one chain.
This makes it possible to talk about the ``integrity of the
 marketplace'' as distinct from the integrity of any one token.

\subsection{Entanglement}
\label{sec:entanglement}
Some \attestations, such as timestamps~\autoref{sec:use-timestamping}, and proofs of work~\autoref{sec:nakamotointegrity}, implicitly lend integrity to everything in a block's ancestry.
When many \addss reference each other's blocks, these recursive \attestations can make some forms of fraud very difficult.
For example, if many applications regularly reference past timestamps,
 and many applications request timestamps from a variety of servers,
 it quickly becomes difficult to falsely claim a block
 \textit{did not} happen before a given time, when doing so would
 involve hiding evidence embedded in many different applications.
\RVR{(was this RVR?): Do we need to cite some Moir thing here?}
\ICS{There is no pre-existing published work that I know of that uses
      this term. 
     This subsection exists in part so we can claim it.
    }

\section{Blockchains as \addss}
\label{sec:blockchainsstructures}
Charlotte is an ideal framework for building new blockchains and
related applications~(\autoref{sec:blockchains1}).
In the simplest sense, a blockchain is any path through the \blockweb.
However, most existing blockchain applications are considerably more
 complicated.

Like all \addss, a blockchain needs \textit{integrity} and
 \textit{availability}.
Here, integrity means that an observer's
 view~(\autoref{sec:observersBeliefsAdversaries})
 always features a \textit{main chain}, in which no two \blocks
 ever have the same \textit{height}.
Availability means that once an observer observes a main chain \block
 at a height, that \block remains available for download indefinitely.

\subsection{Separating Availability and Integrity}
\label{sec:separating}
With few exceptions~\cite{bigchaindb}, existing blockchain systems
 require that all integrity servers (e.g., miners, and consensus
 nodes) store all the blockchain data.
This is fundamentally inefficient.
For example, a traditional byzantine consensus system tolerating $f$
 failures needs $> 3f$ participants, while a storage system tolerating
 $f$ failures needs only $> f$ participants.
If blockchain systems separated storage and consensus duties, they
 would be able to store about 3 times as much as they do, with the
 same failure assumptions.

Charlotte makes it easy to separate availability from integrity.
Wilbur servers store \blocks, and provide
 \availabilityattestations~(\autoref{sec:wilbur}).
References to those \blocks carry those \attestations, proving the \block
 referenced is available. 
Fern servers need only issue \integrityattestations for each \block on
 the chain, rather than storing it themselves.

\ICS{Ethereum example below has been relegated to the tech report, to
      save space.}
\ifreport
For example, if one were to build something like \ethereum in
 Charlotte, what \ethereum calls \textit{\block headers} would
 themselves be \integrityattestations, and the Merkle root in each
 would instead be a reference (or collection of references) to \blocks
 stored on Wilbur servers.
This makes it natural to search and retrieve \block headers and portions
 of state, without splitting apart \blocks, or downloading the whole
 chain.
\fi

\subsection{Integrity Mechanisms}
\label{sec:blockchainintegrity}
Different blockchains have used a variety of mechanisms to maintain
 the integrity of the
 chain~\cite{bitcoin, mazieresstellar, Cachin2017}.
To demonstrate the flexibility of Charlotte, we have implemented a few
 example mechanisms in small-scale experiments.

\subsubsection{Nakamoto (Proof-or-Work)}
\label{sec:nakamotointegrity}
We can represent a \bitcoin / \ethereum style blockchain as an \adds
 $D$ whose states are trees of proof-of-work \blocks.
An observer with security parameter $k$ (say, 6) believes only in
 universes with a main chain that grows faster than any side chains
 differing by $k$ or more \blocks.
More precisely, if a universe $U$ includes a state $S$ of such a
 blockchain $D$ featuring a fork of $k$ or more \blocks, one side of
 the fork must be the \textit{main} chain, and all main chain \blocks
 $k$ or higher above the root of the fork must be observed before
 $\p{\observebefore U}$ all other \blocks in $S$ of equal height.
\ICS{Long sentence}

\subsubsection{Agreement}
\label{sec:agreementintegrity}
Some blockchain applications only require agreement: they lose
 liveness if two \textit{potentially} valid \blocks are proposed for
 the same height~\cite{avalanche, nano, spectre}.
For instance, if a chain represents a single bank account, and
 potentially valid \blocks represent transactions signed by the account
 holder, then honest account holders should never sign two
 transactions unordered by the \blockweb. 

Agreement servers are simple to implement.
When a server attests to a \block, it promises never to attest to any
 conflicting \block.
For a given server, an agreement \attestation type $\tau$ does not
 feature any universes where two conflicting \blocks both have an
 \attestation from the server.
Observers can construct quorums of trusted servers, as
 in~\autoref{sec:integrityattestations}.
A \block appears in an observer's view  when the observer has observed
 enough \attestations: committing a conflicting \block would require
 too many parties to break their promises.

\subsubsection{Heterogeneous Consensus (Hetcons)}
\label{sec:hetcons}
%

is our own consensus
 algorithm based on Leslie Lamport’s Byzantine
 Paxos~\cite{byzantizing-paxos}.
Hetcons allows each pair of observers
 (\textit{learners}~\cite{byzantizing-paxos}) to specify the set of
 universes~(\autoref{sec:observersBeliefsAdversaries}) in which they
 must agree. 
They define these universes in terms of which Fern servers
 (\textit{acceptors}~\cite{byzantizing-paxos}) are \textit{safe} (not
 Byzantine) and \textit{live} (not crashed) in each.
In this way, we support heterogeneous servers, heterogeneous
 (or ``mixed''~\cite{Schneider90}) failure models, and
 heterogeneous observers.
In the symmetric case, when all observers have the same failure
 tolerances, Hetcons reduces to regular Byzantine
 Paxos~\cite{byzantizing-paxos}.

\ifreport
\ICS{The following is relegated to the tech report, because I'm not
      sure it's strictly necessary to understand Hetcons as an example
      integrity mechanism for Charlotte.}
For each observer, Hetcons forms quorums of participants whose
 \attestations are necessary to put a block in the observer's view.
Generally speaking, two observers will agree so long as all their
 quorums intersect on a safe participant.
\fi

\subsection{Blocks on Multiple Chains}
\label{sec:multichain}
\label{sec:meet}
In general, nothing prevents a single \block from being part of
 multiple chains.
It simply requires the \integrityattestations for each chain.
For example, if one blockchain represents records of events that have
 happened to a specific vehicle (crashes, repairs, \dots),
 and another represents repairs a specific vendor has performed, it
 makes sense to append the record of a specific repair to both chains.
The record (a \block) could reference the previous \blocks on
 each chain, and the next \blocks on each would in turn reference it.
Each chain's integrity mechanism would have to attest to
 the \block, and references to the \block could carry both sets of
 \attestations to let readers know it is in both \addss.

\subsubsection{Atomicity}
Sometimes, such \block appends need \textit{atomicity}.
\ICS{Should we switch to the example where you're trying to buy a
      train ticket and a hotel room atomically?}
For example, suppose one blockchain represents the cryptocurrency
 \red{\blockchainName{redCoin}}, and another represents the
 cryptocurrency \blue{\blockchainName{blueCoin}}.
\principalA wants to give \principalB one
 \red{\blockchainName{redCoin}} in exchange for one
 \blue{\blockchainName{blueCoin}}.
This represents two transactions: one on each chain.
Crucially, either \textit{both} happen, or \textit{neither} do. 
Otherwise, it's possible that \principalA will give \principalB a
 \red{\blockchainName{redCoin}}, and get nothing in return.
We want to commit both transactions together, \textit{atomically}.

\subsubsection{Meet}
To atomically commit one block to multiple \addss, we require a single
 \integrityattestation which represents a commitment to all of them.
We call the type of this \integrityattestation the
 \textit{meet}~$\p\meet$ of the types of the \integrityattestations
 for the \addss involved.
If an \attestation of type $\tau_{\red r}$ commits a \block to
 \red{\blockchainName{redCoin}}, and an \attestation of type
 $\tau_{\blue b}$ commits a \block to \blue{\blockchainName{blueCoin}},
 then an \attestation of type $\tau_{\red r} \meet \tau_{\blue b}$
 commits a \block to both.
\ifreport
\ICS{The following is relegated to the tech report, because I'm not
      sure it's strictly necessary to understand the fact that meets
      exist.
     We might want to move it back, because meet fits so nicely into
      our model.
}
In a sense, $\tau_{\red r} \meet \tau_{\blue b}$ is a subtype of
 $\tau_{\red r}$ or $\tau_{\blue b}$, since an \attestation of the meet
 type can be used wherever an \attestation of either supertype can.
In our types-as-observers semantics~(\autoref{sec:availabilityattestations}), we
 define meet as $\meet \triangleq \cap$.
The assumptions made by the meet type encapsulate all the assumptions
 made by its component types.
\fi

Not all pairs of \integrityattestation types have a meet.
However, we created meet types for our Hetcons blockchains.
\ICS{Do we want to relegate the rest of this subsection to the tech report?}
The quorum necessary for an \attestation with the meet
 type is the union of one quorum from each component type.
In other words, to make an observer decide with an 
 \integrityattestation of type $\tau_{\red r}\meet \tau_{\blue b}$,
 you need all the participants it would take to make an \attestation
 of type $\tau_{\red r}$ \textit{and} all the participants it would
 take to make an \attestation of type $\tau_{\blue b}$.
With this construction, we can atomically commit a single \block onto
 multiple Hetcons chains.

\subsection{Linearizable Transactions on Objects}
\label{sec:linearizable}
It can be useful to model state as a collection of stateful
 \textit{objects}, each of which has some availability and integrity
 constraints~\cite{raz92}.
\ICS{better citation desired}
We can model objects as a chain of \blocks, defined by availability
 and \integrityattestations upholding these constraints.
For instance, if an object must be consistent and available so long as
 3 of a specific 4 servers are correct, each \block should have ``store
 forever'' \availabilityattestations from 2 servers, and
 \integrityattestations from 3 stating that they'll never attest to
 any other \block in that slot.

Each \block represents a state change for each of the
 objects represented by chains of which the \block is a part.
In other words, the \blocks are \textit{atomic}
 (or \textit{ACID}) \textit{transactions} in the database
 sense~\cite{Haerder1983}.
\ifreport
\ICS{ The following sentence just isn't a very interesting claim.}
A collection transactions is guaranteed to have a consistent, serial
 order so long as the chains maintained for each of the objects
 they touch are consistent.
\fi
For a given observer, the transactions involving
 objects which that observer assumes to be linearizable have a
 serial order so long as that observer's assumptions are correct.
Furthermore, two correct observers can never see two transactions
 oppositely ordered.

This gives programmers a natural model for atomic transactions across
 object-chains with different integrity and availability mechanisms,
 which would be useful for applications from banking to supply chain
 tracking.
Transactions can involve any set of objects, so long as their
 integrity mechanisms have a meet operation for atomic commitment
 (\autoref{sec:multichain}).
\ICS{I have relegated the example applications to the tech report.
     I am not sure this was the right thing to do.
     However, we need to save space.
}

\ifreport
\subsubsection{Banking}
\label{sec:banking}
We can imagine bank accounts as a linearizable objects, with state
 changes being deposits and withdrawals to and from other bank
 accounts, signed by appropriate parties.
We can model this in Charlotte.
Each bank maintains some integrity mechanism (Fern servers) to ensure
 accounts' state changes are totally ordered, which prevents
 double-spending.
Likewise, each bank maintains some Availability mechanism
 (Wilbur servers), ensuring transactions relevant to their customers'
 accounts aren't forgotten.
Each transaction is thus a \block shared by two chains, and must be
 committed atomically onto both chains.

When considering how ``trustworthy'' the money in an account is, what
 matters is the integrity of the \adds featuring the full
 ancestry of all transactions in the account.
To ensure the trustworthiness of their accounts, banks may issue their
 own \integrityattestations for all transactions in the causal past of
 transactions involving that bank.
This requires checking that ancestry for any inconsistencies with
 anything to which the bank has already attested.
This ensures any observers trusting the bank's attestations have
 consistent view~(\autoref{sec:observersBeliefsAdversaries}), but
 cannot guarantee that observers trusting different banks have the
 same view.

An ``attest to the complete history'' approach is analogous to
 auditing the full finances of everyone with whom you do business for
 every transaction.
In reality, much of the time, banks effectively trust each other's
 \attestations.
This allows much faster transaction times with weaker guarantees.

\subsubsection{Supply Chain Tracking}
Much like bank accounts, we can imagine each good in a supply chain as
 a linearizable object. 
Transactions may involve decreasing / destroying some goods to
 increase / create others.
For example, a transaction might feature destroying 10 kg from a case
 of grapes to add 9 kg to a vat of juice, and 1 kg to a bin of
 compost.
As with banking, each good is only as ``trustworthy'' as the \adds
 featuring its complete ancestry, and audits / attesting to
 past transactions can increase this trustworthiness.
\fi

\subsection{Application to Payment Graphs}
\label{sec:bitcoinanalysis}
\begin{figure}
 	\centering
{
\begin{tabular}{| l   r | l | l |}
\hline
& & \textbf{Unaltered} & \textbf{2 Accounts} \\
\hline
\multirow{2}{*}{linearized}   & {longest chain}      &
\bitcoinnumtransactions & \bitcoinTwoTXTotal
\\\cline{2-4}
                              & time  &
\textbf{\bitcointotaltime} & \bitcoinTwoTXTotalTime
\\\hline
\multirow{2}{*}{parallelized} & {longest chain}      &
\bitcoinlengthlongest & \bitcoinlengthlongestweighted
\\\cline{2-4}
                              & time  &
\bitcoincharlottetotaltime{}& \textbf{\bitcoincharlotteweightedtotaltime}
\\\hline
\end{tabular}
}
 	\caption{Theoretical advantages of Charlotte-style parallelization in the \bitcoin payment network}
 	\label{fig:bitcoinparallel}
\end{figure}

The Charlotte framework makes it easy to imagine parallelized
 blockchain-based payments, with each account as a stateful object,
 represented by a
 chain~(\ifreport\autoref{sec:banking}\else\autoref{sec:linearizable}\fi).
As the \bitcoin payment network is a popular example of blockchain-based
 finance, we consider the theoretical advantages offered by
 parallelization in a Charlotte-style approach.
 
Bitcoin does not keep track of money in terms of accounts.
Instead, each transaction divides all its money into a number of
 outputs, called Unspent Transaction Outputs, or UTXOs,
 each of which specify the conditions under which they can be
 spent (e.g., a signature matching this public key).
Each transaction specifies a set of input UTXOs as well, from which
it gets the money, and it provides for each a proof that it is
 authorized to spend the money (e.g., a digital signature).
Each UTXO is completely drained when it is spent, and cannot be
 reused.
Thus, Bitcoin transactions form a graph, with transactions as
 vertices and UTXOs as directed edges~\cite{bitcoin}.

In our Charlotte banking model, each bank account is a chain, so
 a transfer between two accounts is simply a \block on two
 chains~(\autoref{sec:linearizable}).
Therefore, if two sets of financial transactions don't
 interact, they can operate entirely in parallel.
The speed of the system is limited by the speed of its slowest chain.
If appending a transaction to its chains takes constant time, the
 speed limit is simply the length of the longest chain.

Blocks~\bitcoinstartblock{} through \bitcoinendblock{} of \bitcoin
contain \bitcoinnumtransactions{} transactions.
The longest chain through this graph has length \bitcoinlengthlongest{},
 so in principle, Charlotte needs time for only
 \bitcoinlengthlongest{} rounds of consensus to accommodate the entire
 payment graph.
Although \bitcoin batches several transactions per \block, it
 required \bitcoinnumblocks{} rounds of consensus to do the same,
 taking a total of \bitcointotaltime{}.
Thus, even with a similarly slow consensus mechanism, a parallelized
 Charlotte approach, even with no batching, would require only
 \bitcoincharlottetotaltime{}.
Of course, Charlotte bank accounts can specify Fern servers with
 whatever consensus mechanism they like.
This could be a much faster system, such as PBFT~\cite{pbft}.

In \bitcoin, it improves anonymity and performance to combine many
 small transfers of money into big ones, with many inputs and many
 outputs.
In the real financial system of the USA, however, all monetary
 transfers are from one account to another.
They are all exactly two-chain transactions.
We can simulate this limitation by refactoring each transaction
 as a DAG of transactions with logarithmic depth
 (Appendix~\ref{sec:twoaccounts}).

With this construction, a Charlotte banking system might use more than
 one transaction per \bitcoin transaction.
The longest chain through this new transaction graph has length
 \bitcoinlengthlongestweighted{}; so, in principle, Charlotte
 can process the entire graph in 
only this many rounds of consensus.
Thus, even with a consensus mechanism as slow as that of Bitcoin,
 Charlotte would still require only
 \bitcoincharlotteweightedtotaltime{}, a speedup of 28.


\section{Implementation}
\label{sec:impl}
Our full Charlotte spec, with all example types and APIs, is 298 lines
 of gRPC (mainly protobuf)~\cite{grpc}.
We implemented proof-of-concept servers in 3833 lines of
 Java~\cite{java-11ed} (excluding comments and import statements),
 with a further 1133 lines of unit tests.
We also wrote 1149 additional lines of Java setting up various
 experiments.
\ifblinded
Anonymized code is available~\cite{anonymized-charlotte-code}.
\else
Anonymized code is available~\cite{anonymized-charlotte-code}.
\fi

\ifreport
\begin{figure}
\centering
\begin{lstlisting}[language=protobuf3,style=protobuf]
message PublicKey {
  message EllipticCurveP256 {
    bytes byteString;}
  oneof keyalgorithm_oneof {    
    AnyWithReference any;
    EllipticCurveP256 ellipticCurveP256;}}
message CryptoId {
  oneof idtype_oneof {
    AnyWithReference any;
    PublicKey publicKey;
    Hash hash;}}
message Signature {
  message SHA256WithECDSA {
    bytes byteString;}
  CryptoId cryptoId;
  oneof signaturealgorithm_oneof {
    AnyWithReference any;
    SHA256WithECDSA sha256WithEcdsa;}}
\end{lstlisting}
\caption{Signature Specification.
  We include \texttt{Any} types for extensibility, as well as default
   built-in types, like \texttt{Sha256WithECDSA}.
  Note that the \blue{\texttt{message}} keyword defines a type in the local
   scope.}
\label{fig:signaturespecification}
\end{figure}
\fi

\subsection{Wilbur servers}
\label{sec:wilburimp}
By default, our example Wilbur servers store all \blocks received in
 memory forever.
They are not meant to be optimal, but they are usable for
 proof-of-concept applications.
The only type of \availabilityattestation we have implemented is
one in which the Wilbur servers promise to
 store the \block indefinitely.
This attestation proves that the \block is available as long as
 the Wilbur server is functioning correctly.

Our Wilbur servers can be configured with a list of known
 peers, to whom they will relay any \blocks they receive and any
 \attestations they create.
This is easy to override: servers can be made to relay \blocks to any
 collection of peers.


We also implemented the WilburQuery service of \autoref{sec:wilburquery}.
Our Wilbur servers can do fill-in-the-blank pattern
 matching on all implemented block types.
The Wilbur Query service imposes no overhead on other services.

\subsection{Version Control}
\label{sec:versioncontrolimpl}
\ifreport
\begin{figure}
\centering
\begin{lstlisting}[language=protobuf3,style=protobuf]
message SignedGitSimCommit {
  message GitSimCommit {
    message GitSimParents {
      message GitSimParent {
        Reference parentCommit;
        bytes diff;}
      repeated GitSimParent parent;}
    string comment;
    Hash hash;
    oneof commit_oneof {
      bytes initialCommit;
      GitSimParents parents;}}
  GitSimCommit commit;
  Signature signature;}

message Block {
  oneof blocktype_oneof {
    AnyWithReference any;
    string protobuf;
    SignedGitSimCommit signedGitSimCommit;}}

message IntegrityAttestation {
  message GitSimBranch {
    google.protobuf.Timestamp timestamp;
    string branchName;
    Reference commit;}
  message SignedGitSimBranch {
    GitSimBranch gitSimBranch;
    Signature signature;}
  oneof integrityattestationtype_oneof {
    AnyWithReference any;
    SignedGitSimBranch signedGitSimBranch;}}
\end{lstlisting}
\caption{Git Simulation \integrityattestation Specification.
  We include \texttt{Any} types for extensibility, and provide
   types like \texttt{SignedGitSimBranch} as options.
  Note that the \blue{\texttt{message}} keyword defines a type in the
   local scope, and that the \texttt{Signature} type is defined in the
   full Charlotte spec~\cite{anonymized-charlotte-code}.}
\label{fig:gitsimspecification}
\end{figure}
\fi
We implemented a simulation of Git~\cite{git}.
Our servers are not fully-functional version control software, as they
 do not implement file-diffs and associated checks, which are
 irrelevant for the purpose of demonstrating the Charlotte framework.

\ifreport
The types for our version control \adds are described in
 \autoref{fig:gitsimspecification}.
\fi
We created a block subtype, \texttt{SignedGitCommit}, representing a
 specific state of the files tracked.
Each block features a signature, comment, hash of the state.
It can be an \textit{initial} commit, in which case it has no
 parents, but does include bytes representing the full contents of the
 files being tracked.
Alternatively, it can have some number of parent commits, each with a
 reference and a file diff.

\ICS{I think we've been through some of this material before, but I
      don't know how much anymore}
A Version Control Fern server tracks the current commit it associates
 with each \textit{branch} (strings).
They issue \integrityattestations that declare which commits they've
put on which branches.
A correct Fern server should never issue two
 such \attestations for the same branch,
 unless the commits they reference are ordered by the \blockweb.
In other words, each new commit on a branch should follow from the
 earlier commit on that branch; it cannot be an arbitrary jump to some
 other files.
Our example servers enforce this
 invariant~\cite{anonymized-charlotte-code}.

Fern servers can have other reasons to reject a request to put a
 commit on a branch. 
Perhaps they accept only commits signed by certain keys.
When a client issues a request, they can include \attestation
 references.
A Fern server can demand that clients prove a commit is, for instance,
 stored on certain Wilbur servers before it agrees to put it on a 
 branch.
The Wilbur servers need not even be aware of the Git data types.

Our version control implementation can use the same Wilbur servers as
 any other application.
In fact, separating out the storage duties of Wilbur from the
 branch-maintaining duties of Fern allows our Charlotte-Git system to
 divide up storage duties of large repositories, much like
 git-lfs~\cite{gitlfs}.

\subsection{Timestamping}
\label{sec:timestampingimpl}
\ifreport
\begin{figure}
\centering
\begin{lstlisting}[language=protobuf3,style=protobuf]
message IntegrityAttestation {
  message TimestampedReferences {
    google.protobuf.Timestamp timestamp;
    repeated Reference block;}
  message SignedTimestampedReferences {
    TimestampedReferences timestampedReferences;
    Signature signature;}
  oneof integrityattestationtype_oneof {
    AnyWithReference any;
    SignedTimestampedReferences sigTimeRefs;}}
\end{lstlisting}
\caption{Timestamping \integrityattestation Specification.
  We include \texttt{Any} types for extensibility, and provide
   \texttt{SignedTimestampedReferences} as an option.
  Note that the \blue{\texttt{message}} keyword defines a type in the
   local scope, and that the \texttt{Signature} type is defined in the
   full Charlotte spec~\cite{anonymized-charlotte-code}.}
\label{fig:timestampingspecification}
\end{figure}
\fi
Timestamps are a subtype of \integrityattestation.
Each Timestamp includes a collection of references to earlier \blocks,
 the current clock time~\cite{open-group-base-7},
 and a cryptographic signature.

Our Timestamping Fern servers timestamp any references requested,
 using the native OS clock.
By default, they issue a timestamp immediately for
 any request, and do not need to actually receive the \blocks
 referenced.
Because references contain hashes, the request itself
 guarantees the \block's existence before that time.

Our Timestamping Fern servers also implement \textit{batching}.
Every 100 (configurable at startup) timestamps, the
 Fern server issues a new timestamp, referencing the
 \blocks it has timestamped since the last batch.
Each server then submits its batch timestamp to other Fern servers
(configurable at startup) for timestamping.
Since timestamps are transitive (if $a$ timestamps $b$, and $b$
 references $c$, then $a$ also timestamps $c$), \blocks are very
 quickly timestamped by large numbers of Fern servers.
This allows applications to quickly gather very strong
 timestamp integrity.

\subsection{Blockchains}
\label{sec:blockchainsimpl}
In principle, any path through the \blockweb is a
 blockchain~(\autoref{sec:blockchainsstructures}).
We implemented Fern servers using three very different integrity
 mechanisms~(\autoref{sec:blockchainintegrity}).
We used some of these servers to demonstrate the advantages of
 separating integrity and availability
 mechanisms~(\autoref{sec:separating}), and blockchain composition:
 we put \blocks on multiple chains~(\autoref{sec:multichain}).

\subsubsection{Agreement}
\label{sec:agreementimpl}
\ifreport
\begin{figure}
\centering
\begin{lstlisting}[language=protobuf3,style=protobuf]
message IntegrityAttestation {
  message ChainSlot {
    Reference block;
    Reference root;
    uint64 slot;
    Reference parent;}
  message SignedChainSlot {
    ChainSlot chainSlot;
    Signature signature;}
  oneof integrityattestationtype_oneof {
    AnyWithReference any;
    SignedChainSlot signedChainSlot;}}
message IntegrityPolicy {
 oneof integritypolicytype_oneof
  { AnyWithReference any;
    IntegrityAttestation fillInTheBlank;}}
\end{lstlisting}
\caption{Agreement \integrityattestation Specification.
  We include \texttt{Any} types for extensibility, and provide
   \texttt{SignedChainSlot} as an option.
  Note that the \blue{\texttt{message}} keyword defines a type in the
   local scope, and that the \texttt{Signature} type is defined in the
   full Charlotte spec~\cite{anonymized-charlotte-code}.}
\label{fig:agreementspecification}
\end{figure}
\fi
Our Agreement Fern servers keep track of each a blockchain as a
 \textit{root} \block, and a set of \textit{slots}.
Each slot has a number representing distance from the root of the
 chain.

Our Agreement Fern servers use the \texttt{SignedChainSlot} subtype of
 \integrityattestation\ifreport~(\autoref{fig:agreementspecification}).
It features a cryptographic signature, and references to a chain's
 \textit{root}, a slot number, and the \block in that slot.
This serves as a format for both requests and \attestations.
Each request is simply an
 \texttt{IntegrityAttestation} with some fields (like the
 cryptographic signature) missing.
While it is possible to encode this in the \texttt{IntegrityPolicy}'s
 \texttt{any} field, we provide the \texttt{fillInTheBlank} option as
 a convenience.

The Agreement Fern servers
\else. They
\fi
 are configured with parameters describing
 which requests they can accept, in terms of requirements on the
 reference to the proposed \block and its parent.
Once a correct Agreement Fern server has attested that a \block is in a
 slot, it will \textit{never} attest that a different \block is in that
 slot.
For instance, to configure a blockchain using quorums of 3 Agreement
 Fern to approve each \block, we require that each request's
 \texttt{parent} Reference include 3 appropriate
 \integrityattestations.

Our Agreement Fern servers make it easy to separate integrity and
 Availability duties~(\autoref{sec:separating}).
To ensure that a \block is available before committing it to the chain,
 we require a \texttt{\block} Reference to include specific
 \availabilityattestations from Wilbur servers.

\subsubsection{Nakamoto}
\label{nakamotoimpl}
\ifreport
\begin{figure}
\centering
\begin{lstlisting}[language=protobuf3,style=protobuf]
message IntegrityAttestation {
  message NakamotoIntegrityInfo {
    Reference block;
    Reference parent;}
  message NakamotoIntegrity {
    NakamotoIntegrityInfo info;
    uint64 nonce;}
  oneof integrityattestationtype_oneof {
    AnyWithReference any;
    NakamotoIntegrity nakamotoIntegrity;}}
\end{lstlisting}
\caption{Nakamoto \integrityattestation Specification.
  We include \texttt{Any} types for extensibility, and provide
   \texttt{NakamotoIntegrity} as an option.
  Note that the \blue{\texttt{message}} keyword defines a type in the local
scope.~\cite{anonymized-charlotte-code}.}
\label{fig:nakamotospecification}
\end{figure}
\fi

Nakamoto, or \textit{Proof of Work} Consensus is the integrity
 mechanism securing \bitcoin~\cite{bitcoin}.
We model it formally in \autoref{sec:nakamotointegrity}.
In \bitcoin, \textit{miners} create proofs of work, which are stored
 by \textit{full nodes}.
With the Simplified Payment Verification \ifreport (SPV)\fi protocol,
 \textit{clients} submit a transaction, and retrieve the
 \textit{\block headers} (proofs of work and Merkle roots) of each
 \block in the chain from full nodes~\cite{bitcoin}.
Each client can use these to verify that its transaction is in the
 chain (has integrity).

We implement miners as Fern servers, which produce
 \integrityattestations bearing proofs of work, taking the place of
 \block headers.
Wilbur servers take the place of full nodes, and store \blocks,
 including \integrityattestations.
For simplicity, our implementation assumes one transaction per
 \block, so clients generate \blocks, and request attestations.
When a client receives an
 \integrityattestation\ifreport~(\autoref{fig:nakamotospecification})\fi,
 it can retrieve the full chain from Wilbur servers.

\ifreport
\ICS{I'm not sure it's necessary to include this detail, even in the TR.}
With SPV, Clients traditionally try to collect \block
 headers until they see their transactions buried
 ``sufficiently deep'' in the chain.
For simplicity, our Fern servers delay responding to the client at
 all until the client's \block has reached a specified (configurable)
 depth.
Regardless, clients can collect \integrityattestations from Wilbur
 servers until they're satisfied.
\fi

Our implementation of Nakamoto consensus offers a
 more precise availability guarantee than \bitcoin does.
Nakamoto Fern servers demand \availabilityattestations with any \blocks
 submitted, ensuring that before a block is added to the chain, it
 meets a (configurable) availability requirement.

\subsubsection{Heterogeneous Consensus}
\label{sec:hetconsimpl}
We implemented a prototype of Hetcons~(\autoref{sec:hetcons}) as a
  Fern service.
Integrity attestations are specific to each observer's assumptions.
We use Charlotte blocks as messages in the consensus protocol
 itself, so \attestations can reference messages demonstrating that
 consensus was achieved.

Hetcons inherits Byzantine Paxos' minimum latency of 3 message delays.
In our implementation, clients do not participate in the consensus:
 they merely request an \integrityattestation from a Fern server.
Including receiving a request from and sending an \attestation to the
client, the process has a minimum latency of 5 messages\ifreport~(\autoref{fig:hetconsrtt}).

\begin{figure}
 	\centering
 	\includegraphics[height=0.4\textwidth]{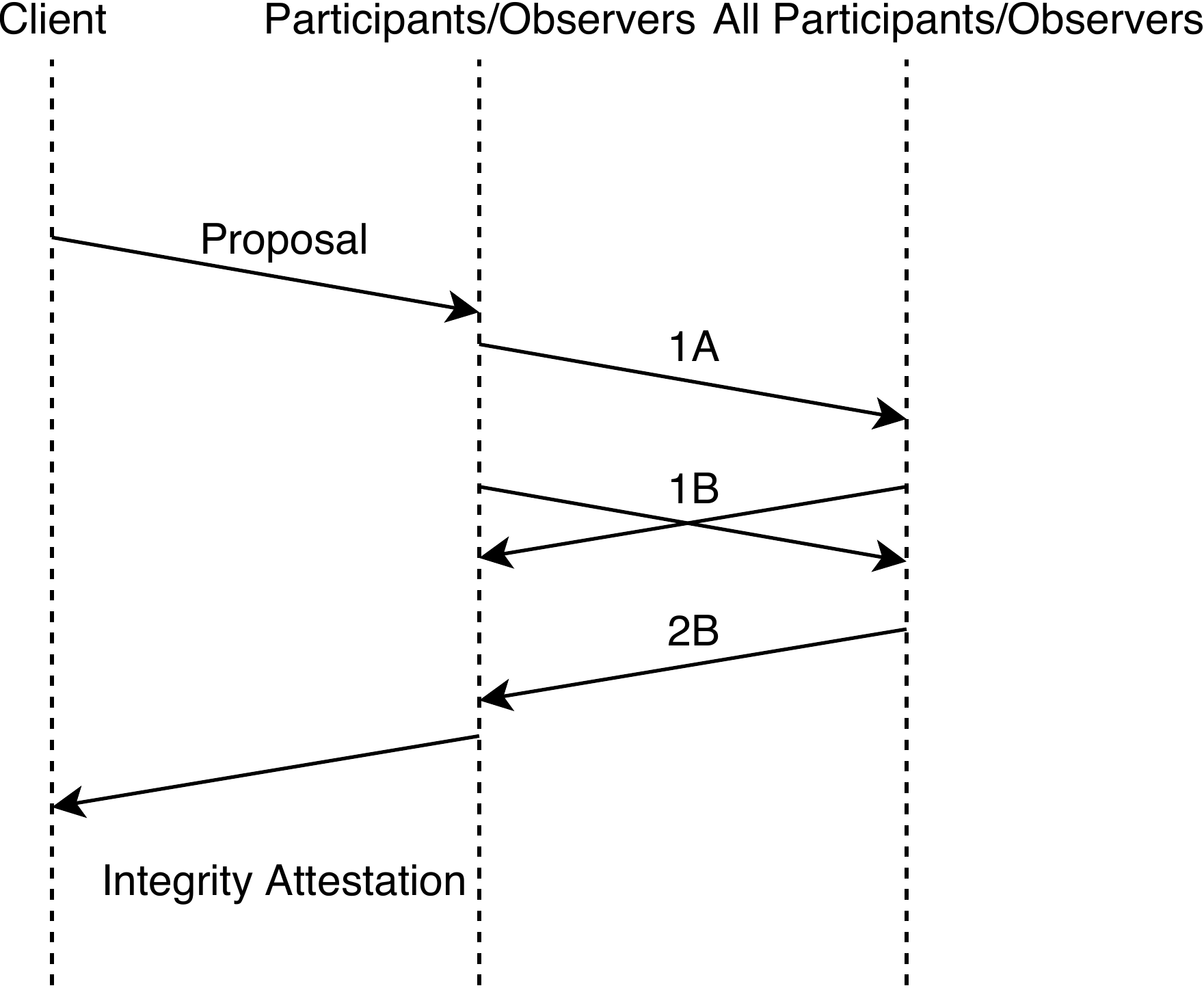}
 	\caption{Heterogeneous Consensus Message Latency}
 	\label{fig:hetconsrtt}
\end{figure}

\else .
\fi

In our implementation, quorums representing trust configurations are
 encoded as \blocks.
Each Hetcons blockchain includes a reference to such a \block in its
 root, ensuring everyone agrees on the configuration.
To append a \block to the chain, a client requests an
 \integrityattestation for some observer, specifying proposed \block
 and height.
To propose one \block be appended to multiple chains, a client can
 request an \integrityattestation that is the meet (\autoref{sec:meet})
 of the \integrityattestations needed for both chains.
The Fern servers then run a round of consensus in which each quorum
 includes one quorum of the consensus necessary for each chain.

For the purposes of demonstrating the Charlotte framework, our
 experiments with Hetcons are \textit{symmetric}: all observers
 want to agree under the same conditions.
For instance, observers might trust 4 Fern servers to maintain a
 chain, expecting no more than one of them to be Byzantine.

\section{Evaluation}
\label{sec:evaluation}

To evaluate the performance of Charlotte, we ran instances of each
 example application.
Except as specified, experiments were run on a local cluster using
 virtual machines with Intel E5-2690 2.9 GHz CPUs, configured as
 follows:

\begin{itemize}
\item Clients: 4 physical cores, 16 GB RAM
\item Wilbur servers: 1 physical core,  8 GB RAM
\item Fern servers: 1 physical core, 4 GB RAM
\end{itemize}

To emulate wide area communication, we introduced 100 milliseconds
 artificial communication latency between VMs.

\subsection{Blockchains}
\label{sec:blockchainseval}

Since blockchains are an obvious application of Charlotte, we
 evaluated the performance, scalability, and compositionality of
 various blockchain implementations.

\subsubsection{Nakamoto}
\label{sec:nakamotoeval}
\begin{figure}
\centering
\includegraphics[width=\ifreport0.9\else0.5\fi\textwidth]{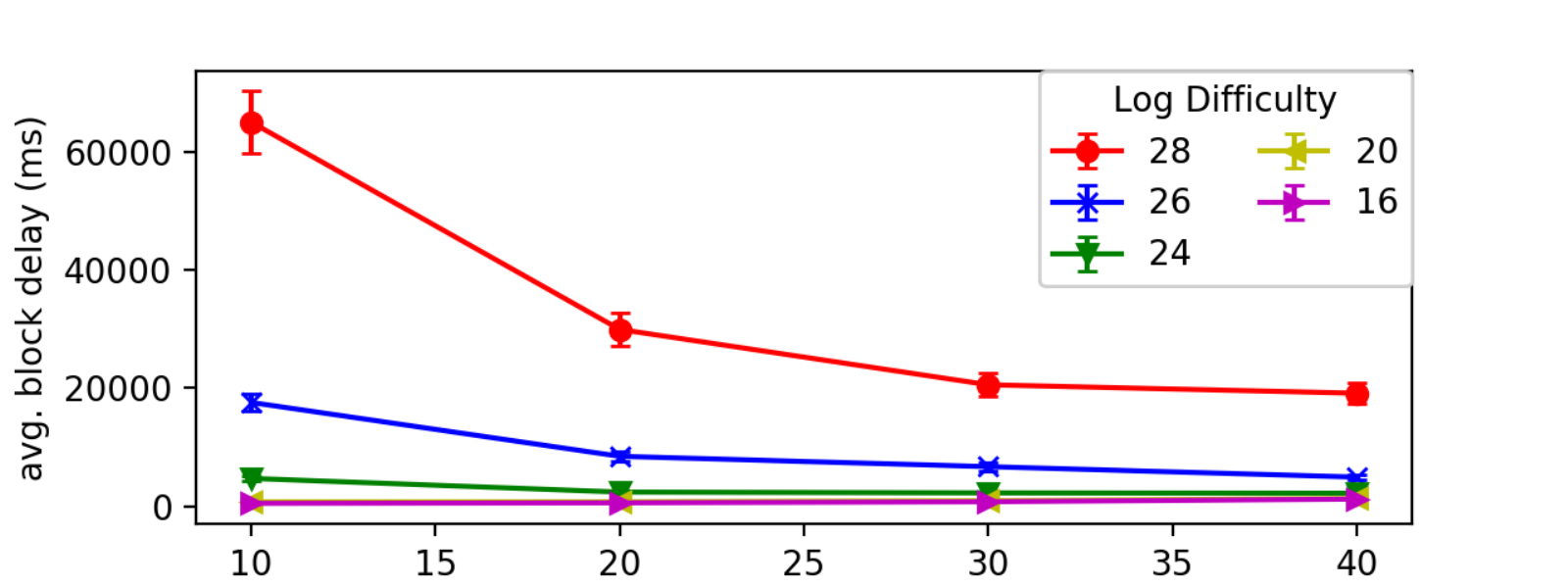}
\caption{Mean \block delay of Nakamoto on Charlotte, with
         bars showing standard error.
         Difficulty is represented in $log_2$ of the number of
          hashes expected to mine a new block.
        }

\label{fig:nakamoto}
\end{figure}

To compare performance of our Nakamoto implementation to
 \bitcoin's, we used multiple ($n=10, 20, 30, 40$)
 Charlotte nodes and measured the mean delay (across 100 consecutive
 blocks) until a client received an \integrityattestation for a \block
 with fixed security parameter $k = 1$.
All clients and servers had one physical core, and 4 GB RAM.
\autoref{fig:nakamoto} shows the results of our tests with various
 difficulty values (expected number of hashes to mine a \block).

When difficulty is low, the delay for an \integrityattestation is
 dominated by the communication overhead (200 ms).
When, more realistically, the difficulty is high, delay is dominated
by the cost of mining.
\ICS{Latency actually increases linearly with difficulty, but
      Log(difficulty) is shown, and linear latency is show, so it
      looks like an exponential.}
\autoref{fig:nakamoto} shows that latency increases with difficulty
 and decreases with the inverse of the number of Charlotte servers
 (total computational power).
Charlotte indeed scales suitably for blockchain
 implementations.

In fact, \bitcoin has about $2 \times 10^{11}$ times the hash
 power~\cite{bitcoinenergy}, and $10^{14}$ times the difficulty as we
 had in our experiment, and it achieves an average \block latency of
 10 min.
With compute power scaled appropriately, our implementation
 would achieve comparable performance: about 5 minutes per \block.

\subsubsection{Agreement}
\label{sec:eval-agreement}
To evaluate the bandwidth advantages of separating integrity
 and availability services, we built Agreement
 Chains~(\autoref{sec:agreementimpl}) tolerating 1--5 Byzantine
 failures, both with and without Wilbur servers.
To tolerate $f$ Byzantine failures, a chain needs $3f+1$ Fern servers,
 and, if it relies on Wilbur servers for availability, $f+1$ Wilbur
 servers.
We tested the latency and bandwidth of our chains, with some
 experiments using 10 byte \blocks, and some using 1 MB \blocks.
In each experiment, a single client appends 1000 blocks to a chain,
 with the first 500 excluded from measurements to avoid
 warm-up effects.
Each experiment ran three times.

In the simple case, without Wilbur servers, all Fern servers receive
 all \blocks. 
This resembles the traditional blockchain strategy~\cite{bitcoin}.
The theoretical minimum latency is 2 
round trips from the client to the Fern servers, or 200 ms.

We also built chains that separate the Fern servers' integrity duties
 from Wilbur servers' availability duties. 
In these chains, Fern servers would not attest to any reference unless
 it included $f+1$ different Wilbur servers'
 \availabilityattestations. 
\ifreport
\else
For brevity, we discuss latency in Appendix~\ref{sec:agreementlatency}.
\fi

\ifreport
\paragraph{Latency}
\begin{figure} \centering
\begin{tabular}{rc}
\rotatebox[origin=l]{90}{\hspace{2cm} milliseconds}&\hspace{-5mm}
\includegraphics[width=8cm]{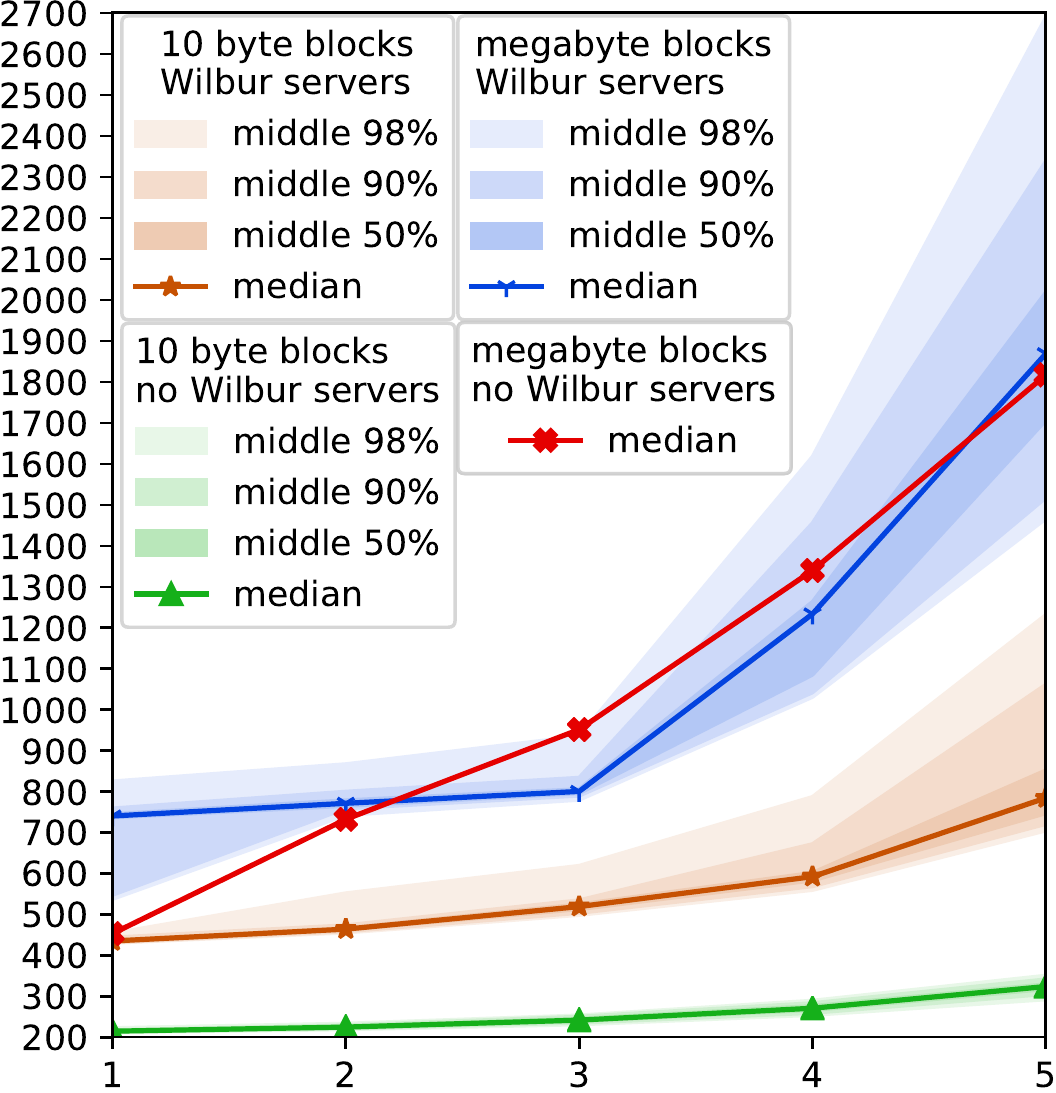}
\\ & Byzantine Failures Tolerated
\end{tabular}
\caption{
Time to commit \blocks in Agreement chains with various numbers of
 servers.
The shaded zones cover the middle percentile of \blocks, so the top of
 the lightest zone represents the 99th (slowest) percentile, and the
 bottom represents the 1st (fastest) percentile. 
The distribution for the megabyte-block,
 no-wilbur-server experiment is
 in~\autoref{fig:agreementpercentile2}.
}
\label{fig:agreementpercentile}
\end{figure}

\begin{figure} \centering
\begin{tabular}{rc}
\rotatebox[origin=l]{90}{\hspace{2cm} milliseconds}&\hspace{-5mm}
\includegraphics[width=8cm]{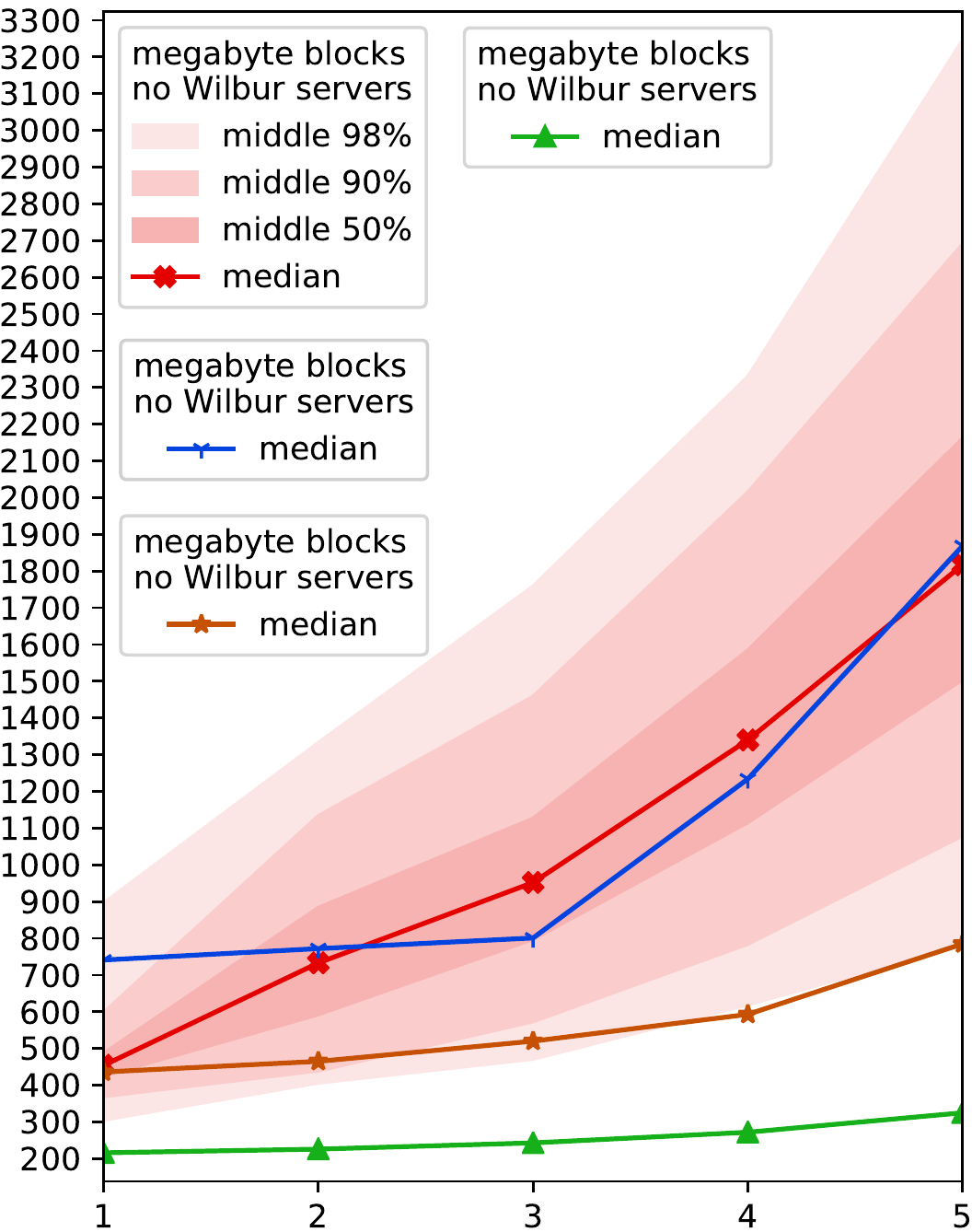}
\\ & Byzantine Failures Tolerated
\end{tabular}
\caption{Time to commit \blocks in Agreement chains with various numbers of servers. The distribution for the megabyte-block, no-wilbur-server experiment is shown.}
\label{fig:agreementpercentile2}
\end{figure}

\autoref{fig:agreementpercentile} and
 \autoref{fig:agreementpercentile2}  show the median latency to commit
 a \block for each of our Agreement chain
 experiments\ifreport\else~(\autoref{sec:eval-agreement})\fi.
Theoretical minimum latency is 4 message sends (round trip from the
 client to the Wilbur servers, and then from the client to the Fern
 servers), or 400 ms.
For chains with small \blocks, latency remains close to the 200 ms and
 400 ms minimums.
For chains with 1 megabyte \blocks, experimental setup has significant
 slowdowns, likely due to bandwidth limitations.

%

\fi

\begin{figure}
\centering
\begin{tabular}{rc}
\rotatebox[origin=l]{90}{\hspace{2cm} gigabytes}&\hspace{-5mm}
\ifreport
\includegraphics[width=8cm]{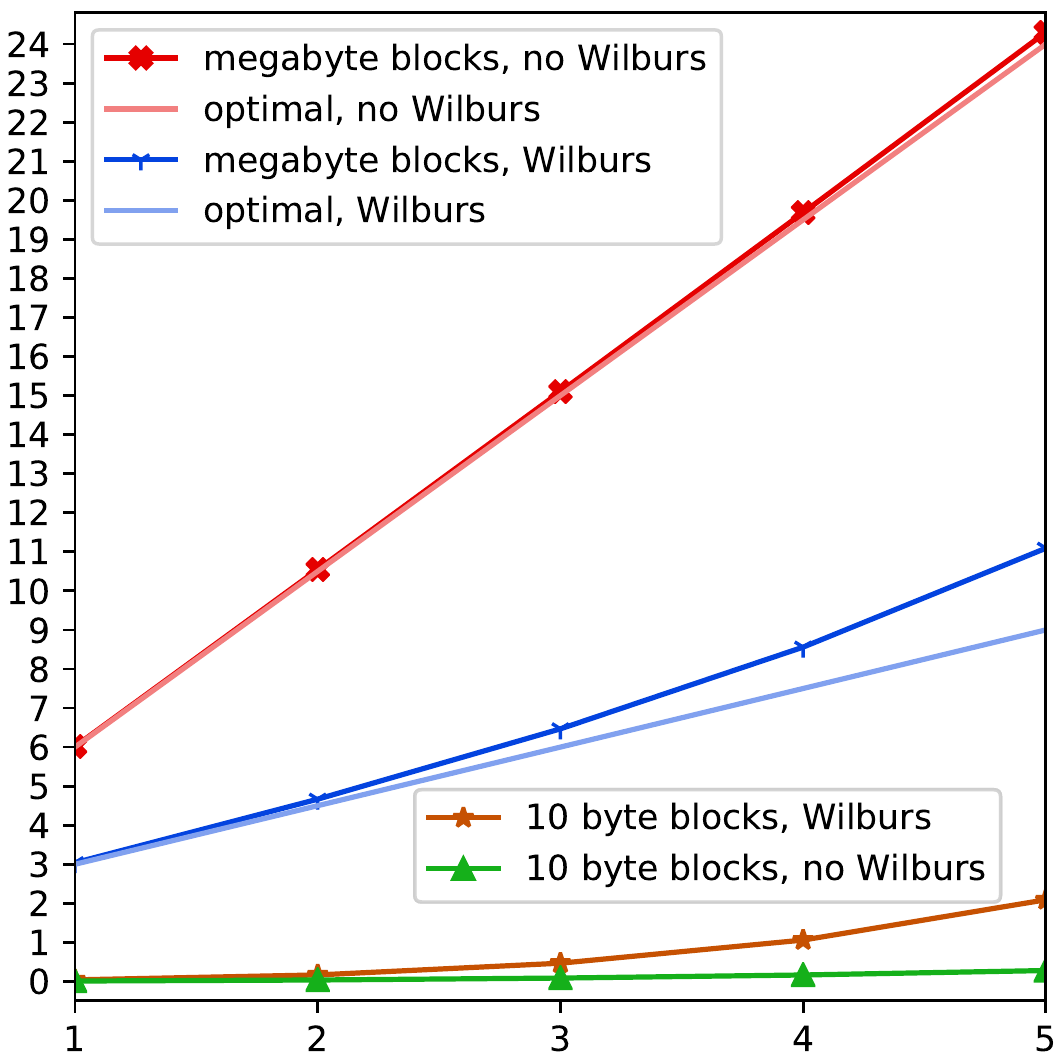}
\else
\includegraphics[width=8cm]{figures/agreement/agreement-bandwidth-short.pdf}
\fi
\\ & Byzantine Failures Tolerated
\end{tabular}
\caption{Total bandwidth used by a client appending 1500 \blocks to Agreement-based chains.}
\label{fig:agreementbandwidth}
\end{figure}

\paragraph{Bandwidth}
Separating availability and integrity
 concerns~(\autoref{sec:separating}) has clear benefits in terms of
 bandwidth.
Because it sends large \blocks to just $f+1$ Wilbur servers
 instead of $3f+1$ Fern servers, our client uses much less bandwidth
 in the large-\block experiments with Wilbur servers than without
 them~(\autoref{fig:agreementbandwidth}).
In theory, committing a \block with Wilbur servers requires bandwidth
for $f+1$ \blocks, and without Wilbur servers requires
 $3f+1$ \blocks.
 The overhead inherent in the additional communication
 with Wilbur servers and the \attestations issued is small
 compared to the savings.

\subsubsection{Heterogeneous Consensus}
\label{sec:hetconseval}
In order to evaluate the feasibility of consensus-based blockchains,
 and multi-chain blocks in Charlotte, we built several chains with
 Hetcons~(\autoref{sec:hetcons}), and ran \ifreport 5 \else 4 \fi
 types of experiments.
With our artificial network latency, the theoretical lower bound on
 consensus latency is 500 ms, and maximum throughput per chain is
 2 \blocks/second.
Each experiment recorded the latency clients experience in
 appending their own \blocks to the chain, as well as system-wide
 throughput.
All Hetcons experiments used single-core VMs with 8 GB RAM, except as noted.

\ifreport
\ICS{Relegated to the tech report, because it's a special case of Parallel.}
\paragraph{Single Chain}
In these experiments, a client appends 2000 successive \blocks to one
 chain.
Mean latency is 527$ms$ for a chain with 4 Fern servers and 538$ms$
 for 7 Fern servers.
Since the best possible is 500 ms, these results are
 promising.
Overheads include cryptographic signatures, verification, and garbage
 collection.
\fi

\begin{figure} \centering
\begin{tabular}{rc}
\rotatebox[origin=l]{90}{\hspace{1cm} Latency(ms)}&\hspace{-5mm}
\includegraphics[width=\ifreport0.9\else0.4\fi\textwidth]{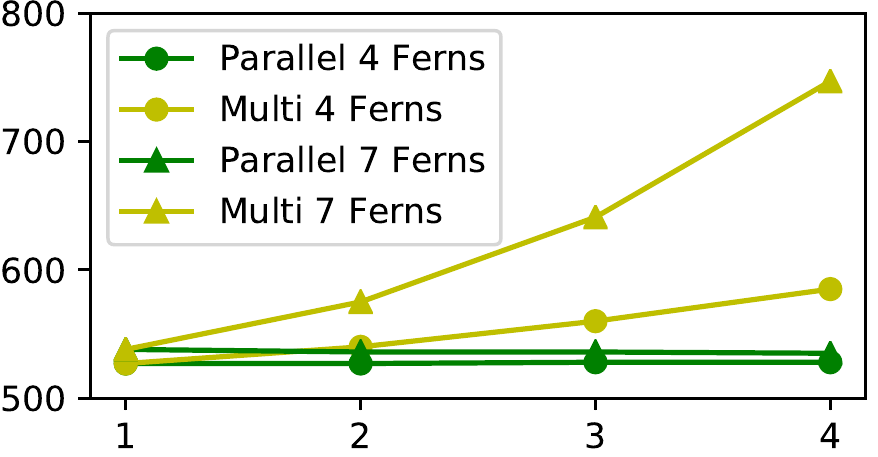}
\\ & Number of Chains
\end{tabular}
\caption{Hetcons Multichain and Parallel experiments.
In Parallel experiments, each chain operates independently
 (and has its own client).
In Multichain experiments, one client tries to append all blocks to
 all chains.
Optimal latency is 500 ms.} 
\label{fig:hetconsManyMulti}
\end{figure}

\paragraph{Parallel}
As the darker green lines in \autoref{fig:hetconsManyMulti} show,
 independent Hetcons chains have independent performance.
In these experiments, we simultaneously ran 1--4 independent chains,
 each with 4 or 7 Fern servers.
In each experiment, a client appends 2000 successive \blocks to one
 chain.
There is no noticeable latency difference between a single chain and
 many chains running together.
Throughput scales with the number of chains
 (and inversely to latency).
This scalability is the fundamental advantage of a
 \blockweb over forcing everything onto one central blockchain.
   
\paragraph{Multichain shared \blocks}
Shared (joint) \blocks facilitate inter-chain
 interaction~(\autoref{sec:multichain}).
In these experiments, a single client appends 1000 shared \blocks to
 2--4 chains, each with 4 or 7 Fern servers.
As the yellow lines in \autoref{fig:hetconsManyMulti} show,
 latency scales roughly linearly with the number of chains.

\begin{figure} \centering
\begin{tabular}{rc}
\rotatebox[origin=l]{90}{\hspace{0cm} Throughput(blocks/sec)}&\hspace{-5mm}
\includegraphics[width=\ifreport0.9\else0.4\fi\textwidth]{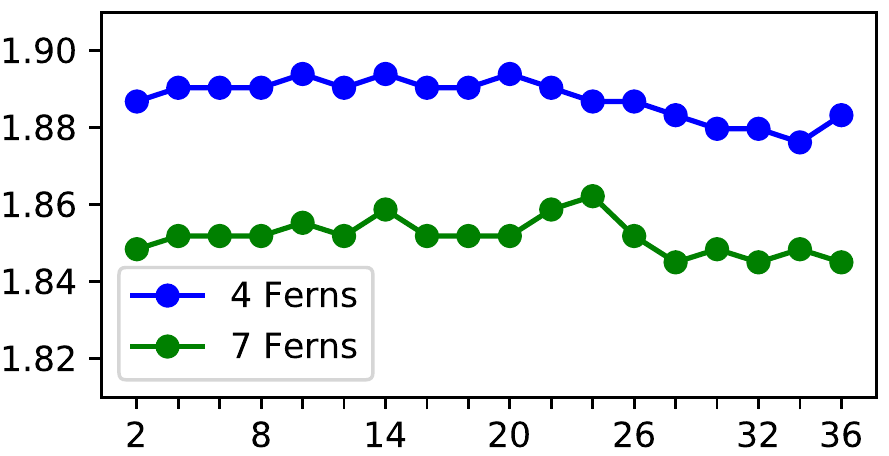}
\\ & Number of Clients
\end{tabular}
\caption{Throughput of Hetcons under contention. 2--36 clients try to
          append 2000 blocks to just one chain. Optimal throughput is 2
          \blocks/sec.}
\label{fig:hetconsContentionThroughput}
\end{figure}

\paragraph{Contention}
In these experiments, all clients simultaneously contend to append
 2000 unique \blocks to the same chain.
We measured the blocks that were actually accepted into slots
 500--1500 of the chain.
We used 2--36 clients, and chains with 4 or 7 Fern servers,
configured with 2 GB RAM.
Like Byzantized Paxos~\cite{byzantizing-paxos}, Hetcons can get stuck
 under contention and occasionally requires a dynamic timeout to
 automatically trigger a new round.
Chain throughput is shown in
 \autoref{fig:hetconsContentionThroughput}. 
Our chains, on average, achieved 1.88 \blocks/sec throughput for 4
 Fern servers and 1.85 \blocks/sec for 7, not far from the
 2 \blocks/sec optimum.
\ACM{Are the results really accurate enough that we can report
 throughput to hundredths? What's the standard error?}
Throughput does not decrease much with the number of clients.


\begin{figure} \centering
\begin{tabular}{rc}
\rotatebox[origin=l]{90}{\hspace{0.2cm} Throughput(blocks/sec)}&\hspace{-5mm}
\includegraphics[width=\ifreport0.9\else0.4\fi\textwidth]{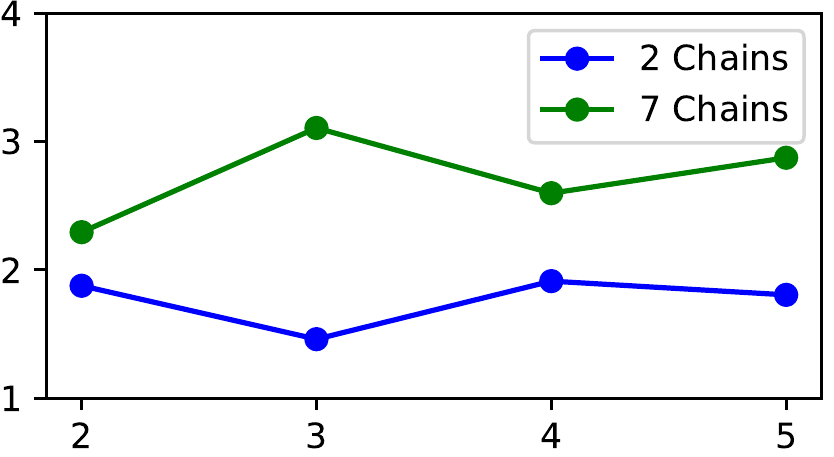}
\\ & Number of Clients
\end{tabular}
\caption{Throughput of Hetcons mixed-workload experiment (4 Fern
servers).}
\label{fig:hetconsMixedThroughput} 
\end{figure}

\paragraph{Mixed} 
These experiments attempt to simulate a more realistic scenario by
 including all 3 types of workload.
2--5 clients contend to append blocks onto either 2 or 7 chains, each
 with 4 Fern servers.
On each block, a client tries to append a shared block to two random
chains with probability 10\% and otherwise tries appending to a
random single chain.
The results are in \autoref{fig:hetconsMixedThroughput}.
Throughput can be over 2.0 blocks/sec because multiple clients can
 append blocks to different chains in parallel.
Mean throughput is 1.8 blocks/sec and 2.7 blocks/sec for
 2 and 7 chains respectively, which
is expected because the 2-chain configuration has more contention.

\ifreport
\smallskip
Hetcons scales well horizontally with multiple chains running in
 parallel.
Furthermore, throughput does not decrease much with more clients
 involved.

This gives us ability to make progress even with lots of clients
connecting to the same chain concurrently. We also notice that, the
number of Ferns servers play major roles for the latency performance.
With a small group of Fern servers, Hetcons can almost reach 500 ms,
which is the best we can get. Although the latency increases linearly
with respect to the number of Ferns, for some applications, it is possible
to break down big shared \blocks into
a set of small shared \blocks. For example, in 
\autoref{sec:bitcoinanalysis}, we discuss how to break a $k$-chain
transaction into a $\log k$-depth graph whose nodes are small two-chain
transactions. By following the same strategy, the latency would
be reduced to $t \times \log k$, where $t$ is the average latency for
completing a 2-chain \block. Since our Hetcons implementation is just
a prototype, we believe that with
further efforts in optimization, average latency performance can be
improved.
\fi

\subsection{Timestamping}
\label{sec:timestampingeval}

To evaluate performance, compositionality, and
 entanglement~(\autoref{sec:entanglement}) with a non-blockchain
 application, we ran experiments with varying numbers of Timestamping
 Fern servers~(\autoref{sec:timestampingimpl}).
All client and server VMs had 4 GB RAM.
For each experiment, a single client requested timestamps for a total
 of 100,000 \blocks.
For each \block, it requested a timestamp from one server, rotating
 through the Fern servers.

\ICS{This repeats content from sec:timestampingimpl , not sure how to
      avoid that repetition}
For each 100 timestamps a Fern server issued, it would create a new
 \block referencing those 100 timestamps, and request that all other
 Fern servers timestamp this \block.
Since timestamps are transitive (if $c$ is a timestamp referencing
 $b$, and $b$ references $a$, then $c$ also timestamps $a$),
 every \block was soon timestamped by all Fern servers.

To explore Charlotte's compositionality, we also composed our
 (1- or 2-failure-tolerant) Agreement chains with our Timestamping Fern
 servers. 
We saw no statistically significant change in
 chain performance: the overhead of Timestamping
was unmeasurably small.
Each \block was timestamped quickly by directly
 requested Timestamping servers, but entanglement~(\autoref{sec:timestampingeval})
 was limited by the chain rate.

\begin{figure}
\centering
\begin{tabular}{rc}
\rotatebox[origin=l]{90}{\hspace{1cm} milliseconds}&\hspace{-5mm}
\includegraphics[width=\ifreport0.9\textwidth\else8cm\fi]{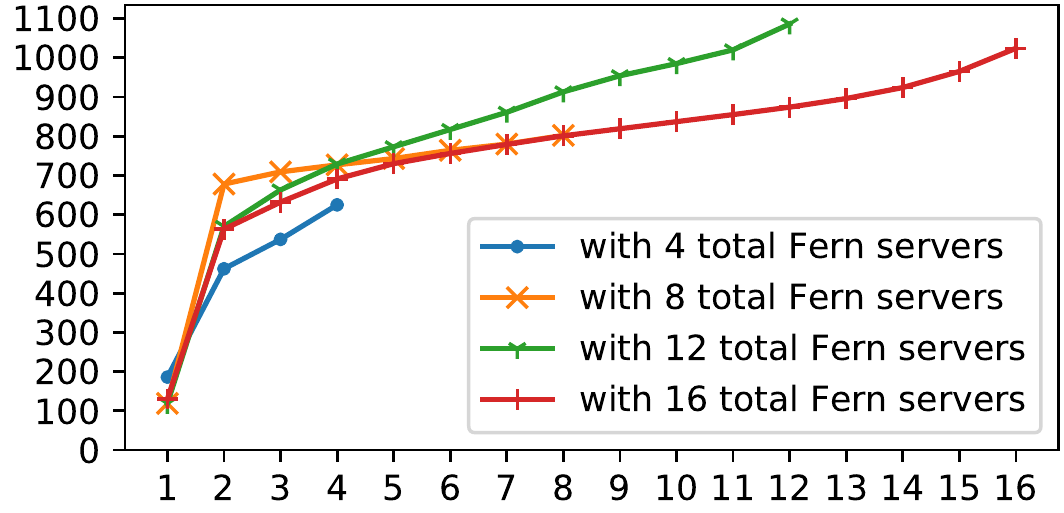}
\\ & Fern servers
\end{tabular}
\caption{Mean time for a \block to be timestamped by $x$ Fern servers, in experiments featuring 4, 8, 12, and 16 total Fern servers.}
\label{fig:timestampmean}
\end{figure}

We also calculated the time it took blocks to
 accrue different Fern servers' timestamps.
As~\autoref{fig:timestampmean} shows, the Fern servers quickly
 timestamp each request.
Blocks get 1 timestamp very close to the 100 ms network latency
 minimum.
There is a delay between 1 and 2 timestamps because
it takes a little while for the Fern servers to collect 100
 timestamps and to create their own \block. 
After that, \blocks accrue timestamps very quickly, since each Fern
 Server requests timestamps from all other Fern servers.
\ifreport
\else
For brevity, we discuss the distribution of block times in
 Appendix~\ref{sec:timestampingpercentile}.
\fi
These experiments suggest that
  entanglement~(\autoref{sec:entanglement}) can be a fast,
  efficient, and compositional way to lend integrity to large \addss. 

\ifreport

\begin{figure}
\centering
\begin{tabular}{rc}
\rotatebox[origin=l]{90}{\hspace{2cm} milliseconds}&\hspace{-5mm}
\includegraphics[width=8cm]{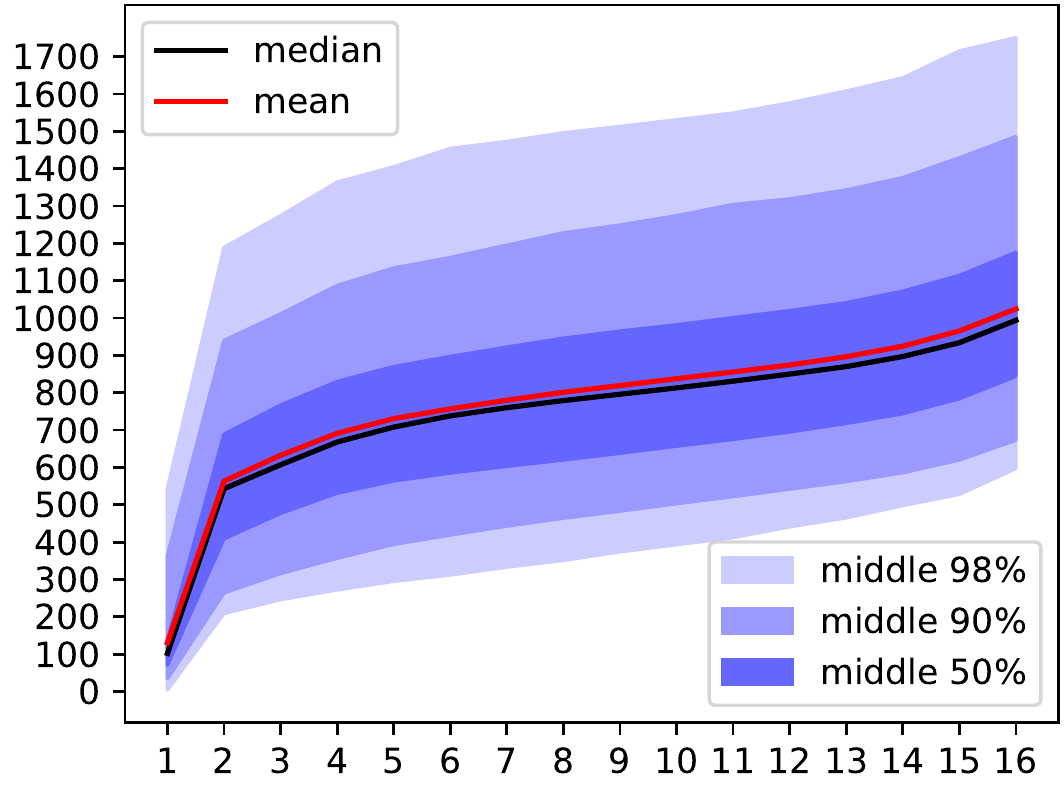}
\\ & Fern Servers
\end{tabular}
\caption{Time for a \block to be timestamped by $x$ Fern Servers, in an experiment featuring a total of 16 Fern Servers.
Shaded zones cover the middle percentile of \blocks, so the top of the lightest zone represents the 99th (slowest) percentile, and the bottom represents the 1st (fastest) percentile. 
Also shown are mean and median \block times
 (very similar). 
}
\label{fig:timestamppercentile}
\end{figure}

Not all \blocks took exactly the same amount of time to accrue the
 same number of timestamps. 
\autoref{fig:timestamppercentile} shows the distribution of times for
 \blocks in the experiment with 16 Fern Servers.
The scale is the same as in~\autoref{fig:timestampmean}. 
In general, each data point (time for \blocks to accrue $x$ timestamps
 in an experiment with $n$ Fern Servers) was approximately Poisson
 distributed. 

\fi

\section{Related Work}
\label{sec:related}

\subsection{Address by Hash}
\label{sec:relatedaddress}
Many other distributed systems reference content by hash, forming
 \addss.
Most reference schemes, however, only work within a specific
 application.
For instance, git uses hashes to reference and request commits stored
 on a server~\cite{git}.
Git-lfs can track and request large files on separate servers with
 hash-based identifiers~\cite{gitlfs}.
Similarly, PKI systems~(\autoref{sec:pki}) reference keys and
 certificates by hash, and maintain groups of availability
 servers~\cite{x509, ldappgp, Kulynych2018}.
Distributed Hash Tables, such as CFS~\cite{cfs} ultimately maintain
 Availability servers, and ensure integrity by referencing data via
 Hash.

HTML pages can reference resources using the \texttt{integrity}
 field~\cite{subresourceintegrity} to specify a hash, and the
 \texttt{src} field to specify a server, like an
 \availabilityattestation without formal guarantees.
Likewise, BitTorrent's Torrent files~\cite{bittorrent} and
 Magnet URIs~\cite{magneturi} reference a file by hashes of
 various kinds, and can specify ``acceptable sources'' from which to
 download the file.
Charlotte's references aim to be extensible in
 terms of the hash algorithms used, and generic over all types of
 data.
Uniquely, Charlotte bundles references to data with
 references to \attestations, which can offer precise
 formal guarantees.

In concurrent work, Protocol Labs' IPLD~\cite{ipld} is a
 multi-protocol format for addressing arbitrary content by hash.
Like Charlotte's \texttt{AnyWithReference}~(\autoref{sec:api}), Multiformats~\cite{multiformats} offers
 an extensible format for
 self-describing data including protobufs~\cite{protobufs}.
Both IPLD and Multiformats are developed closely with
 IPFS~\cite{ipfs}, a peer-to-peer file distribution system.
IPLD references do not include \attestation references the
 way Charlotte references do, but future work could
 fruitfully combine these technologies with
 Charlotte's reference and \block encoding formats.

\subsection{BlockDAGs}
\label{sec:relateddag}
Other projects have explored DAGs of \blocks in a
 blockchain context.
Many, such as Iota~\cite{iota},
 Nano (also known as RaiBlocks)~\cite{nano},
 Avalanche~\cite{avalanche},
 Spectre~\cite{spectre},
 Phantom, and Ghostdag~\cite{phantom}
 are tailored to cryptocurrency.
Each defines its own currency, and they do not compose.

Some projects, such as æternity~\cite{aeternity},
 alephium~\cite{alephium},
 Qubic~\cite{qubic},
 and Plasma~\cite{plasma} enable
 general-purpose computation on a BlockDAG by way of smart contracts.
However, they ultimately rely on a single global consensus mechanism
 for the integrity of every application.

Sharded blockchains, including Omniledger~\cite{omniledger},
 Elastico~\cite{Luu2016},
 RapidChain~\cite{rapidchain},
 RSCoin~\cite{rscoin},
 and Ethereum 2.0~\cite{ethereum-sharding-phase-1}
 are a form of BlockDAG.
Most still require that all applications have essentially the same
 trust assumptions.

Other sharded blockchain projects, such as Aion~\cite{aion},
 Cosmos~\cite{cosmos}, and Polkadot~\cite{polkadot}, envision
 heterogeneous chains with inter-chain communication.
Polkadot features a single Relay Chain trusted by all parachains
(parallelizable chains),
 although it does allow parachains to proxy for
 outside entities, including other blockchains.
Perhaps most similarly to our multi-chain
 transactions~(\autoref{sec:multichain}), Aion can use
 Bridges, consensus mechanisms trusted by multiple chains,
 to commit a transaction to each.

All of these blockchain projects operate at a higher level of
 abstraction than Charlotte.
Charlotte is a generic format for communicating \blocks, with a novel
 \attestation-based model for specifying availability and integrity
 properties. However,
we believe any of these projects could benefit from building their
 implementations within the Charlotte framework.
For example, where 
 Cosmos' Inter-Blockchain Communication~\cite{cosmos} and
 Aion's Transwarp Conduits~\cite{transwarp}
 require that one chain be able to
 read and validate transaction commits from another, we present a
 unified framework for the data they must request and interpret:
 \integrityattestations.

\subsection{Availability \attestations}
\label{sec:relatedavailability}
Although storage services are widely
 available~\cite{amazonstorage,googlestorage,microsoftstorage},
 \availabilityattestations~(\autoref{sec:availabilityattestations})
 make Wilbur servers unique.
The only type of \availabilityattestation we have implemented is a
 simple promise to store a \block indefinitely. 
However, there is a great deal of work on
 reliable storage~\cite{Dimakis2011, Garay97}
 and proofs of retrievability~\cite{Juels2007, Bowers2009, Shi2013}
 that could be used to make a variety of \availabilityattestation
 subtypes that provide more availability with less trust.

\subsection{Integrity \attestations}
\label{sec:relatedintegrity}
Integrity \attestations abstract over a variety of mechanisms lending
 integrity to data provenance and \adds properties.
In some ways, \attestations resemble the labels of distributed
 information flow control systems~\cite{dstar,fabric09}, and
 implement a kind of endorsement~\cite{zznm02}
 as additional \attestations are minted for the same \block.
In other ways, \integrityattestations generalize ordering services for
 traditional distributed systems~\cite{Hunt2010} or
 blockchains~\cite{Sousa2018}.
These services maintain a specific property of a \adds
 (ordering), much like our blockchain \integrityattestations.
However, \integrityattestations generalize over many possible
 properties: timestamps, provenance, etc.

Future \integrityattestation subtypes might take advantage of
 technologies like authentication logic proofs and artifacts representing
 assurances of data provenance~\cite{af99, sirer11}.

\section{Conclusion}
\label{sec:conclusion}
Charlotte offers a decentralized
 framework for composable \Adds
 with well-defined availability and integrity properties.
Together, these structures form the \blockweb, a novel
 generalization of blockchains.
Charlotte addresses many of the shortcomings of existing \addss by
 enabling parallelism and composability.
Charlotte is flexible enough to enable applications patterned after
 any existing \adds while offering rigorous guarantees
 through \attestations that can be given precise semantics.

\ifacknowledgments
\section{Acknowledgments}


\fi

\finalpage

{\footnotesize \bibliographystyle{acm}
\bibliography{bibtex/pm-master,charlotte}}

\appendix

\ifreport
\section{Bitcoin Transactions in Two Accounts or Fewer}
\else

\subsection{gRPC Specifications}
\label{sec:specificationsappendix}
\subsubsection{Charlotte API~(\autoref{sec:api})}
\begin{figure}
\centering
\begin{lstlisting}[language=protobuf3,style=protobuf]
message AnyWithReference {
 google.protobuf.Any any;
 Reference typeBlock;}
message Hash {
 oneof hashalgorithm_oneof
   { AnyWithReference any;
     bytes sha3; }}//technically unnecessary
message Reference {
 Hash hash;
 repeated Hash availabilityAttestations;
 repeated Reference integrityAttestations;}
message Block {
 oneof blocktype_oneof
  { AnyWithReference any;
    string protobuf; }}
\end{lstlisting}
\caption{Core Types of Charlotte: this (slightly simplified) proto3 code describes how \blocks, references to \blocks, and generic data are safely marshaled and unmarshaled in Charlotte. }
\label{fig:types}
\end{figure}

We use Protocol Buffer (protobuf) syntax to describe the
 Charlotte protocols.
\autoref{fig:types} presents the  core types used by Charlotte
 protocols, using Protocol Buffer syntax.
For simplicity, our specifications omit the indices of the various
 fields.
The actual source code is also slightly more complicated for
 extensibility~\cite{anonymized-charlotte-code}.

\begin{figure}
\centering
\begin{lstlisting}[language=protobuf3,style=protobuf]
message SendBlocksResponse {
 string errorMessage;}
service CharlotteNode {
 rpc SendBlocks(stream Block)
     returns (stream SendBlocksResponse) {}}
\end{lstlisting}
\caption{All Charlotte servers implement the Charlotte\-Node service.}
\label{fig:charlottenodeservice}
\end{figure}

In our API, all Charlotte servers must implement the
 \texttt{SendBlocks} RPC (\autoref{fig:charlottenodeservice}),
 which takes in a stream of \blocks and can return a stream of
 responses that may contain error messages.
We define subtypes of \attestation for \textit{Availability} and
 \textit{Integrity}, and show how to construct and observer from
 quorums of types they trust~(\autoref{sec:availabilityattestations}
 and \autoref{sec:integrityattestations}).

\begin{figure}
\centering
\begin{lstlisting}[language=protobuf3,style=protobuf]
message AvailabilityPolicy {
 oneof availabilitypolicytype_oneof {
  AnyWithReference any; }
}
message RequestAttestationResponse {
 string errorMessage;
 Reference reference;
}
service Wilbur {
 rpc RequestAvailabilityAttestation(
      AvailabilityPolicy)
     returns (RequestAttestationResponse){}
}
\end{lstlisting}
\caption{Wilbur Service Specification.}
\label{fig:wilburservice}
\end{figure}

\paragraph{Wilbur} servers host \blocks,
 providing \textit{availability}.
\ICS{Removed the bit actually describing at a high level what Wilbur 
      servers do, since we've said that multiple times before.
     Should I bring it back?}

In blockchain terminology~\cite{bitcoin}, Wilbur servers correspond to
 ``full nodes,'' which store \blocks on the chain.
In more traditional data store terminology, Wilbur servers are
 key--value stores for immutable data.
The Charlotte framework is intended to be used for building both kinds
 of systems.

In our API, Wilbur servers are Charlotte servers that include the
 \texttt{RequestAvailabilityAttestation}
 RPC~(\autoref{fig:wilburservice}), which accepts a description of the
 desired \attestation, and returns either an error message, or a
 reference to a relevant \availabilityattestation.

\begin{figure}
\centering
\begin{lstlisting}[language=protobuf3,style=protobuf]
message WilburQueryInput {
 oneof wilburquery_oneof {
  Reference reference = 1;
  Block fillInTheBlank = 2; }}
message WilburQueryResponse {
 string errorMessage = 1;
 repeated Block block = 2; }

service WilburQuery {
  rpc WilburQuery(WilburQueryInput)
      returns (WilburQueryResponse) {}}
\end{lstlisting}
\caption{WilburQuery Specification}
\label{fig:wilburqueryservice}
\end{figure}

\paragraph{WilburQuery} servers~(\autoref{sec:wilburquery})
additionally implement the \texttt{WilburQuery} RPC~(\autoref{fig:wilburqueryservice}).

\begin{figure}
\centering
\begin{lstlisting}[language=protobuf3,style=protobuf]
message IntegrityPolicy {
 oneof integritypolicytype_oneof
  { AnyWithReference any; }
}
service Fern {
 rpc RequestIntegrityAttestation(
      IntegrityPolicy)
     returns (RequestAttestationResponse){}
}
\end{lstlisting}
\caption{Fern Service Specification.}
\label{fig:fernservice}
\end{figure}

\paragraph{Fern} servers issue
 \textit{\integrityattestations}, which define the set of \blocks in a
 given \adds.
Among other things, \integrityattestations can be proofs-of-work, or
 records demonstrating some kind of consensus has been reached.
One simple type of \integrityattestation, found in our prototype, is a
 signed pledge not to attest to any other \block as belonging in a
 specific slot in an \adds.
Fern servers generalize ordering or consensus services.
In blockchain terminology~\cite{bitcoin}, Fern servers correspond to
 ``miners,'' which select the \blocks belonging on the chain.

In our API, Fern servers are Charlotte servers that include the
 \texttt{RequestIntegrityAttestation} RPC~(\autoref{fig:fernservice}),
 which accepts a description of the desired \attestation, and returns
 either an error message or a reference to a relevant
 \integrityattestation.

\ICS{On Matthew's advice, I've relegated this section to the tech
 report / appendix.
It consists of us saying we don't do a thing that we never claim to
 do, and then an implementation detail that most readers don't care
 about.}
\subsection{Implementation Limitations of Attestations}
\label{sec:limitationsofattestations}
Since programmers can define their own subtypes of integrity or
 \availabilityattestations, nothing prevents them from encoding
 availability guarantees in an \integrityattestation, or violating the
 \availabilityattestation monotonicity
 requirement~(\autoref{sec:availabilityattestations}).
Programmers who violate the system assumptions naturally lose 
 guarantees.

\ICS{It would be nice to condense this paragraph, but I'm not sure how.}
In our implementation, the only operational distinction between an
 \availabilityattestation and an \integrityattestation is in the
 \texttt{Reference} object.
When one \block references another, it can also reference relevant
 integrity and \availabilityattestations. 
However, whereas an included reference to an \integrityattestation is
 itself a \texttt{Reference} object, an included reference to an
 \availabilityattestation carries only a \texttt{Hash}.
This is because an \integrityattestation might need an
 \availabilityattestations to describe where to obtain the
 \integrityattestation.
However, the same is not true of an \availabilityattestation: it is
 pointless to send \availabilityattestation $b$ just
 to describe where to fetch \availabilityattestation $a$, since it is
 just as easy to send \availabilityattestation $a$ in the first place.

\subsection{Latency in Agreement Blockchains~(\autoref{sec:eval-agreement})}
\label{sec:agreementlatency}

\subsection{Time Distribution in Timestamping Experiments~(\autoref{sec:timestampingeval})}
\label{sec:timestampingpercentile}

\begin{figure}
\centering
\begin{tabular}{rc}
\rotatebox[origin=l]{90}{\hspace{2cm} milliseconds}&\hspace{-5mm}
\includegraphics[width=8cm]{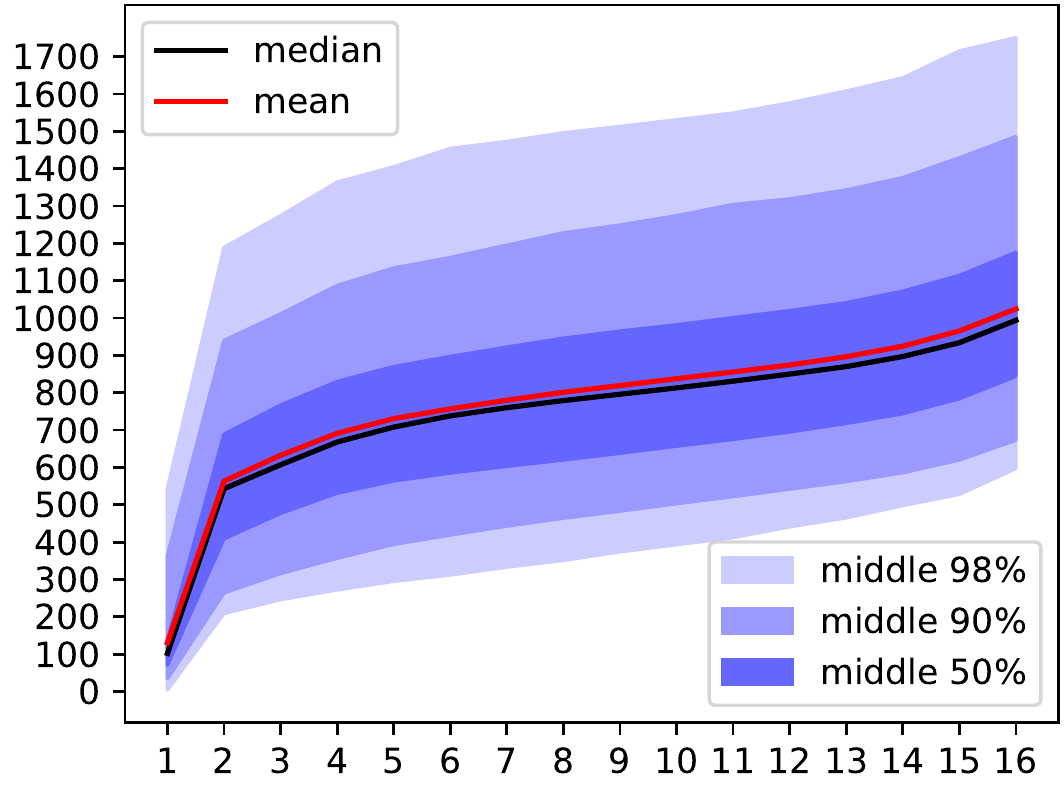}
\\ & Fern Servers
\end{tabular}
\caption{Time for a \block to be timestamped by $x$ Fern Servers, in an experiment featuring a total of 16 Fern Servers.
Shaded zones cover the middle percentile of \blocks, so the top of the lightest zone represents the 99th (slowest) percentile, and the bottom represents the 1st (fastest) percentile. 
Also shown are mean and median \block times
 (very similar). 
}
\label{fig:timestamppercentile}
\end{figure}

Not all \blocks took exactly the same amount of time to accrue the
 same number of timestamps. 
\autoref{fig:timestamppercentile} shows the distribution of times for
 \blocks in the experiment with 16 Fern Servers.
The scale is the same as in~\autoref{fig:timestampmean}. 
In general, each data point (time for \blocks to accrue $x$ timestamps
 in an experiment with $n$ Fern Servers) was approximately Poisson
 distributed.

\subsection{Bitcoin Transactions in Two Accounts or Fewer}
\fi

\label{sec:twoaccounts}
\begin{figure}
  \centering
\begin{tikzpicture}
\draw (.75,1.75) node[draw, circle, text width=5mm, text height = 5mm] (central) {};

\draw[->, line width=0.3mm] node[anchor=north] at (0,.5) {$i_0$} (0,.5) -- (central);
\draw[->, line width=0.3mm] node[anchor=north] at (.5,.5) {$i_1$} (.5,.5) -- (central);
\draw[->, line width=0.3mm] node[anchor=north] at (1,.5) {$i_2$} (1,.5) -- (central);
\draw[->, line width=0.3mm] node[anchor=north] at (1.5,.5) {$i_3$} (1.5,.5) -- (central);

\draw[->, line width=0.3mm] (central) -- (0,3) node[anchor=south] at (0,3) {$o_0$};
\draw[->, line width=0.3mm] (central) -- (.5,3) node[anchor=south] at (.5,3) {$o_1$};
\draw[->, line width=0.3mm] (central) -- (1,3) node[anchor=south] at (1,3) {$o_2$};
\draw[->, line width=0.3mm] (central) -- (1.5,3) node[anchor=south] at (1.5,3) {$o_3$};

\draw[->, line width=2mm] (1.7,1.75) -- (2.4, 1.75);

\draw (3,1) node[draw, circle] (c00) {};
\draw (3,2) node[draw, circle] (c01) {};
\draw (3,3) node[] (c02) {$o_0$};

\draw (4,1) node[draw, circle] (c10) {};
\draw (4,2) node[draw, circle] (c11) {};
\draw (4,3) node[] (c12) {$o_1$};

\draw (5,1) node[draw, circle] (c20) {};
\draw (5,2) node[draw, circle] (c21) {};
\draw (5,3) node[] (c22) {$o_2$};

\draw (6,1) node[draw, circle] (c30) {};
\draw (6,2) node[draw, circle] (c31) {};
\draw (6,3) node[] (c32) {$o_3$};

\draw[->, line width=0.3mm] (2.8,0) -- (c00);
\draw[->, line width=0.3mm] node[anchor=north] at (3,.5) {$i_0$};
\draw[->, line width=0.3mm] (3.2,0) -- (c00);

\draw[->, line width=0.3mm] (3.8,0) -- (c10);
\draw[->, line width=0.3mm] node[anchor=north] at (4,.5) {$i_1$};
\draw[->, line width=0.3mm] (4.2,0) -- (c10);

\draw[->, line width=0.3mm] (4.8,0) -- (c20);
\draw[->, line width=0.3mm] node[anchor=north] at (5,.5) {$i_2$};
\draw[->, line width=0.3mm] (5.2,0) -- (c20);

\draw[->, line width=0.3mm] (5.8,0) -- (c30);
\draw[->, line width=0.3mm] node[anchor=north] at (6,.5) {$i_3$};
\draw[->, line width=0.3mm] (6.2,0) -- (c30);


\draw[->, line width=0.3mm] (c00) -- (c01);
\draw[->, line width=0.3mm] (c00) -- (c11);

\draw[->, line width=0.3mm] (c10) -- (c11);
\draw[->, line width=0.3mm] (c10) -- (c21);

\draw[->, line width=0.3mm] (c20) -- (c21);
\draw[->, line width=0.3mm] (c20) -- (c31);

\draw[->, line width=0.3mm] (c30) -- (c31);
\draw[->, line width=0.3mm] (c30) -- (c01);

\draw[->, line width=0.3mm] (c01) -- (c02);
\draw[->, line width=0.3mm] (c01) -- (c22);

\draw[->, line width=0.3mm] (c11) -- (c12);
\draw[->, line width=0.3mm] (c11) -- (c32);

\draw[->, line width=0.3mm] (c21) -- (c22);
\draw[->, line width=0.3mm] (c21) -- (c02);

\draw[->, line width=0.3mm] (c31) -- (c32);
\draw[->, line width=0.3mm] (c31) -- (c12);
\end{tikzpicture}
  \caption{Converting 4 inputs and 4 outputs to a graph of 2-account transactions.}
  \label{fig:twoaccounts}
\end{figure}
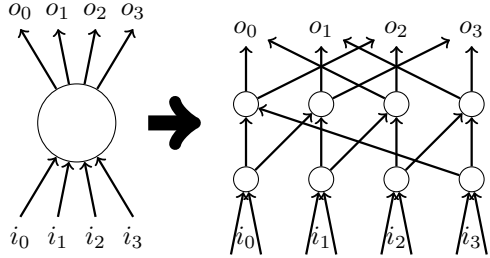

In {\bitcoin}, it is advantageous to combine many small transfers of
 money into big ones, with many inputs and many outputs.
 \ACM{Not sure this appendix is actually worth the space it takes up}
 \ICSreply{Well, we get 5 pages of references and appendices...}
This improves anonymity and performance.
In the real financial system of the USA, however, all monetary
 transfers are from one account to another.
They are all exactly two chain transactions.

We can simulate this limitation by refactoring each {\bitcoin} UTXO as
 2 UTXOs, and each {\bitcoin} transaction as a DAG of transactions with
 depth:
\[
\left\lceil log_2\p{max\p{\textrm{number of inputs}, \textrm{ number of outputs}}} \right\rceil
\]

To do this, we create 
\[
  n \triangleq 2^d
\]
 chains, each of which is
\[
  d \triangleq \left\lceil log_2\p{max\p{\textrm{number of inputs}, \textrm{ number of outputs}}} \right\rceil
\]
 long.
We call these chains $C^0$ through $C^n$.
Original input UTXO $i$ corresponds to both inputs to the first
 transaction of chain $i$.
Original output UTXO $j$ corresponds to one output of each of the last
 transactions from chains $j$ and $\p{j + 2^{d-1}}\textrm{mod }n$.
For $0 \leq k < \p{d-1}$, the outputs of the $k^{th}$ transaction in chain
 $i$, called $C_k^i$, go to $C_{k+1}^i$, and:
\[
  C_{k+1}^{\p{i + 2^j} \textrm{mod}\ n}
\]
The outputs of $C_d^i$ go to the UTXOs corresponding with output $i$,
 and output $\p{i + 2^{d-1}}\textrm{mod }n$.
Each transaction divides its output values proportionately to the sums
 of the final output values reachable from each of the transaction's
 outputs.
\autoref{fig:twoaccounts} is an example transformation from a 4-input,
 4-output transaction to a DAG of depth 2 using all 2-input, 2-output
transactions.

\IfFileExists{tr.ent}
  {\theendnotes}
  {\relax}
\end{document}